\documentclass[submission, Phys]{SciPost}
\usepackage{amsmath}
\usepackage{amssymb}

\usepackage{epstopdf}
\usepackage{MnSymbol}
\usepackage{graphicx}
\usepackage{epsfig}
\usepackage{dcolumn}
\usepackage{bm}
\usepackage{color}
\usepackage{titlesec}
\usepackage{verbatim}

\begin{document}

\begin{center}{\Large \textbf{
Quantum bounds and fluctuation-dissipation relations
}}\end{center}

\begin{center}
Silvia~Pappalardi\textsuperscript{1*},
Laura~Foini\textsuperscript{2},
Jorge~Kurchan\textsuperscript{1}
\end{center}

\begin{center}
{\bf 1} Laboratoire de Physique de l’\'Ecole Normale Sup\'erieure, ENS, Universit\'e PSL, CNRS, Sorbonne Universit\'e, Universit\'e de Paris, F-75005 Paris, France
\\
{\bf 2} IPhT, CNRS, CEA, Universit\'{e} Paris Saclay, 91191 Gif-sur-Yvette, France
\\
* silvia.pappalardi@phys.ens.fr
\end{center}

\begin{center}
\today
\end{center}

\section*{Abstract}
{\bf
In recent years, there has been intense attention on the constraints posed by quantum mechanics on the dynamics of the correlation at low temperatures, triggered by the postulation and derivation of quantum bounds on the transport coefficients or on the chaos rate.
However, the physical meaning and the mechanism enforcing such bounds is still an open question.
Here, we discuss the quantum fluctuation-dissipation theorem (the KMS conditions) as the principle underlying bounds on correlation time scales. By restating the problem in a replicated space, we show that the quantum bound to chaos is a direct consequence of the KMS condition, as applied to a particular pair of two-time correlation and response functions. Encouraged by this, we describe how quantum fluctuation-dissipation relations act in general as a blurring of the time-dependence of correlations, which can imply bounds on their decay rates. Thinking in terms of fluctuation-dissipation opens a direct connection between bounds and other thermodynamic properties.
 }

\vspace{10pt}
\noindent\rule{\textwidth}{1pt}
\tableofcontents\thispagestyle{fancy}
\noindent\rule{\textwidth}{1pt}
\vspace{10pt}

\section{Introduction}
For many years, there has been the intuition that quantum mechanics poses constraints on transport coefficients  such as conductivity \cite{Gunnarsson2003Colloquium, mottiofferegel, Bruin2013Similarity} or viscosity \cite{1985,kovtun}, effective at low temperatures.  This expectation is compatible with the existence of a bound for the physical timescales $\tau$ of a many-body system
\begin{equation}
    \label{boundT}
    \frac 1 \tau \lesssim \frac T \hbar \ ,
\end{equation}          
the so-called Planckian scale determined only by the temperature $T$ and the Planck constant $\hbar$  \cite{Zaanen2004Why, hartnoll2021planckian} (in our units $k_\text B = 1$).
This discussion received an indirect boost from the ``quantum bound to chaos'' proved by Maldacena, Shenker and Stanford \cite{Maldacena2016bound} in 2015. 
They considered a regularized out of time-order correlator (OTOC) and showed that, if there exist a small parameters $\epsilon$, such that the OTOC depends exponentially in time as
\begin{equation}
\label{eq:Malda}
\frac 1Z {\mbox{Tr}} \left( A(t)e^{-\frac{\beta}{4} H}B(0)e^{-\frac{\beta}{4} H}A(t)e^{-\frac{\beta}{4} H}B(0) 
e^{-\frac{\beta}{4} H}\right) \sim a - b \epsilon {e^{\lambda t}}\ ,
\end{equation}
then $\lambda$ shall obey
\begin{equation}
\label{boundL}
\lambda \leq \frac{2 \pi T}{\hbar}  \ .
\end{equation}
The rate $\lambda$ was named quantum Lyapunov exponent since Eq.\eqref{eq:Malda} encodes the classical Lyapunov exponent in the classical limit, as introduced originally by Larkin and Ovchinnikov \cite{larkin1969quasiclassical}.

 The intriguing fact about these quantum bounds is that they are precisely saturated by toy models of holography including the Sachdev-Ye-Kitaev (SYK) model \cite{sachdev, KitaTalk, chowdhury2021sachdev}. As such, these findings resulted in a large body of works in the past few years,  ranging from transport and condensed matter theory to string and quantum field theory up to quantum information theory. There are several outstanding open questions regarding such quantum constraints. What is the relation between the bound to chaos and the transport ones? Are these bounds indicative of the same deep principle in quantum mechanics?
 More recently, Tsuji, Shitara and Ueda provided a rederivation of Eq.\eqref{boundL} that stresses the connection with the quantum Fluctuation-Dissipation Theorem (FDT) or Kubo-Martin Schwinger (KMS) relations \cite{ueda1,ueda2}. See also Ref.\cite{Murthy2019Bounds} for a different approach based on the Eigenstate-Thermalization Hypothesis (ETH) and Ref.\cite{parker2019universal} for a relation between $\lambda$ and the properties of two-point functions, both based on analysis in the frequency domain. 
 Nevertheless, the relation between these two types of bounds, their physical meaning and the underlying mechanism enforcing them is still a matter of ongoing research.

This work contributes to establishing the quantum fluctuation-dissipation theorem as the physical mechanism governing the bounds. The FDT is the cornerstone of statistical mechanics, as it expresses a relation between the intrinsic fluctuations of a system and its linear response to external perturbations. 
Building on the quantum FDT, this paper contains two main results. 1) We map OTOC quantities, as Eq.\eqref{eq:Malda}, into two-time functions in a replicated space, in thermal equilibrium at twice the temperature. As a consequence, this mapping immediately allows to re-derive the Tsuji et al. argument as a simple consequence of the usual quantum-FDT relation governing two-time correlators.
2) We discuss the quantum FDT in time-domain  --- the $t$-FDT --- and we show how it may be interpreted as a blurring of the fine time-details of correlations on an intrinsic timescale
\begin{equation}
    \label{Omega}
  \frac{1}{\tau_{\Omega}} =  \Omega = \frac{\pi T}{\hbar} \ .
\end{equation}
 Under certain conditions, that we discuss, we illustrate that this blurring can result in a transition between dominating timescales that leads to bounds on the decay/growth rate of the response function or intrinsic fluctuations.

The paper is organized as follows.
We begin by recalling the definition of standard two-point functions and the quantum FDT in frequency. Then, we write a quantum time-domain fluctuation-dissipation theorem and we discuss its interpretation as a blurring. 
Next, we describe our first main result: the mapping of OTOC two-times correlation functions in the replicated space and we show a simple derivation of the bound in Eq.\eqref{boundL}. 
The subsequent section contains our second result: a thorough discussion of the effect of the $t$-FDT on correlation functions that depend exponentially on time and the conditions under which it leads to bounds on the decay/growth rate of correlation functions. 
We conclude with a discussion of our findings and open questions.

\section{Correlation and response functions: the quantum FDT}
\label{sec:1}

Given a system at thermal equilibrium with Hamiltonian $H$ and two observables $A$ and $B$, one may define the following two-times correlation functions  \cite{Fabrizio}
\begin{align}
	\label{eq:Rsa}
	S_{AB}(t) & = 	\frac 1Z\text{Tr} \left [ {e^{-{\beta}  H} }\, A(t)      \,  B\,  	\right] 
	= 	C_{AB}(t) +  \hbar R_{AB}''(t) \ ,
	\end{align}
with $C_{AB}(t)$ and $R_{AB}''(t)$ related to the standard fluctuations and response as:
\begin{subequations}
\label{eq:CRdef}
\begin{align}
	\label{eq:Csa}
	C_{AB}(t) & = \frac 12 
	\frac 1Z \text{Tr} \left [ {e^{-\beta  H} }\,\{  A(t) , \,  B\,  \}	\right ]  \ ,
 \\
	R_{AB}(t)  & = {2}i \theta(t) R_{AB}''(t) =
	\frac i \hbar \theta(t) 	\frac 1Z\text{Tr} \left [ {e^{-\beta  H} }\, [ A(t) ,\,  B\,  ]	\right ] \ .
	\end{align}
\end{subequations}
We also define the regulated function: 
\begin{align}
	\label{eq:Fsa}
	F_{AB}(t) & =	\frac 1Z\text{Tr} \left [ {e^{-\frac{\beta}{2}  H} }\, A(t)    {e^{-\frac{\beta}{2}  H} }     \,  B\,  	\right]\ .
	\end{align}
These are often presented in their  Fourier representations: $
 R_{AB}(\omega) = \int_{-\infty}^{\infty}  dt \; e^{  i \omega t} R_{AB}(t)  $ and similarly for all the others. (Time and frequency domain functions will be henceforth distinguished from the arguments).
 The function $F_{AA}(\omega)$ has a clear interpretation in terms ETH, i.e. $F_{AA}(\omega) = |f(E, \omega)|^2$ where $f$ is the smooth function appearing in the ETH ansatz for the matrix elements of operators in the energy eigenbasis $|A_{nm}|^2  \propto |f(E, \omega)|^2$ with $E=(e_n+e_m)/2$, $\omega=e_n -e_m$ and energies $e_n,e_m$ corresponding to temperature $T$ (see \cite{Foini2019Eigenstate}).

The quantum fluctuation-dissipation theorem, the KMS conditions, are easily derived using the Lehman representation \cite{Fabrizio} or the cyclic property of the trace, and read:
\begin{subequations}
    \label{eq:fdt}
\begin{align}
\label{eq:fdtC}
	C_{AB}(\omega) & = \cosh(\beta \hbar \omega /2)  F_{AB}(\omega) 
		\ ,\\
\label{eq:fdtR}
	\hbar R_{AB}''(\omega) & = \sinh(\beta \hbar \omega /2)  F_{AB}(\omega)
		\ ,
\end{align}
\end{subequations}
equivalent to the standard formulation $\hbar  R''_{AB}(\omega) =  \tanh(\beta \hbar \omega /2) C_{AB}(\omega)$. Restricting to self-correlations, namely $A=B$, one has $R''_{AA}(\omega) =  \text{Im} R_{AA}(\omega)$.
Note that $F_{AB}(\omega)$ has to decay sufficiently fast at large frequency in order for $C_{AB}(t)$ and $R_{AB}(t)$ to be defined \cite{Murthy2019Bounds}.
The Matsubara frequencies $i n \Omega$, which represent the zeros of the hyperbolic functions in (\ref{eq:fdtC}) and (\ref{eq:fdtR}), define characteristic timescales which will naturally appear in our analysis in real-time and will correspond to the bounds that we discuss.
Let us also note that if one looks at the expressions of $F_{AB}(\omega)$ in terms of $C_{AB}(\omega)$ or $R''_{AB}(\omega)$ it is clear that the high frequency signal in the correlation or in the response function is strongly suppressed due to the hyperbolic functions. As we now discuss, this effect can be interpreted as a smoothing of the physical quantities in the time domain.

\section{$t$-FDT: from frequency to time}
\label{sec:tFDT}
\begin{figure}[t]
\begin{center}
\includegraphics[width=9cm]{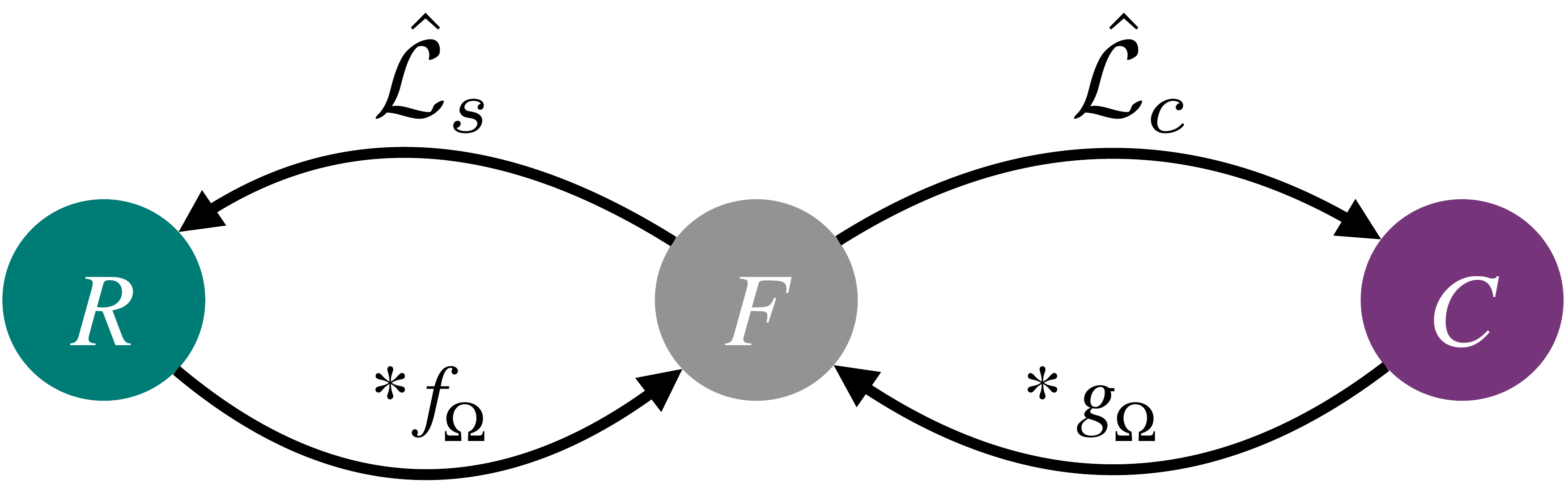}
\caption{Graphical summary of the $t$-FDT.
The differential operators $\hat{\mathcal L}_{c/s}$ act on the regulated fluctuations $F(t)$ leading to $C(t)$ and $R(t)$. Conversely, by convolution with $g_\Omega(t)$ (and $f_\Omega(t)$) one blurs out the details of $C(t)$ (and $R(t)$) going into $F(t)$.}
\label{fig:tFDT}
\end{center}
\end{figure}

 While the quantum FDT has a simple formulation in the frequency
domain, in order to discuss bounds to the decay or the growth of $C$, $R$ and $F$, it will turn out to be much more illuminating to study its formulation in the time domain. This will allow us to deduce physical consequences without resorting to the structure in Fourier space.
The functions $C_{AB}(t)$ and $R_{AB}(t)$ can be written in terms of $F_{AB}(t)$ by application of a differential operator, i.e.
\begin{subequations}
\label{eq:FDTC}
\begin{align}
	\label{eq:FDTCt}
	C_{AB}(t)  & =
	 \frac 12 \left [F_{AB}\left(t+\frac {i\beta \hbar}2\right)+ F_{AB}\left(t-\frac {i\beta \hbar}2\right) \right ] 
	   \equiv \hat{\cal{L}}_c  F_{AB}(t) 
		= \cos \left (\frac {\beta \hbar}2 \frac{d}{dt} \right) F_{AB}(t) 
 	\\
	\label{eq:FDTRt}
	R_{AB}(t) 
	 & 
	  = - \frac{2}{\hbar} \theta(t) \frac{1}{2i}  
	  \left [ F_{AB}\left(t+\frac {i\beta \hbar}2\right)-
	 F_{AB}\left(t-\frac {i\beta \hbar}2\right)\right] 
	\equiv
	 - \frac{2}{\hbar} \theta(t)  \hat{\cal{L}}_s  F_{AB}(t)
	 = - \frac{2}{\hbar} \theta(t) \sin \left (\frac {\beta \hbar}2 \frac{d}{dt} \right) F_{AB}(t)  \ ,
\end{align}	
\end{subequations}
where we have defined $\hat{\cal{L}}_{c/s}$ that act as shifts of time in the imaginary axis. 
Equations (\ref{eq:FDTC}) can be inverted and their consequences have a clear physical interpretation.
In terms of the  \emph{integrated response} 
\begin{equation}\label{Eq_psi}
\Psi_{AB}(t) = \int_0^t {\rm d} t' \; R_{AB}(t')\ ,
\end{equation}
and the \emph{static susceptibility} $\Psi_{AB}(\infty)$, starting from the standard $\omega$-FDT, one finds
\begin{subequations}
		\label{eq:invetFDT}
\begin{align}
	\label{eq:invetFDT_C}
	F_{AB}(t) & =  \int_{-\infty}^{\infty}  C_{AB}(t')\,   g_{\Omega}(t-t')\, {\rm d} t'
	\\
	F_{AB}(t) 
	 \label{eq:invetFDT_R}
	&  =  \int_{-\infty}^{\infty} {\rm d} t' \, iR''_{AB}(t')\, f_{\Omega}(t-t') \ ,
	\\
	& =
	\frac {\Psi_{AB}(\infty) + \Psi_{AB}(\infty)}{2 \beta }
	-
	\frac {1}{2\beta} 
 \int_{0}^{\infty} {\rm d} t' \, 
	\Big [\Psi_{AB}(t')
	 \,  \tilde f_{\Omega}(t'-t) + 
	\Psi_{BA}(t')\, \tilde f_{\Omega}(t'+t) 
	\, \Big ]\ . \nonumber		
\end{align}	
\end{subequations}
The functions $g_\Omega(t)$ and $ f_\Omega(t)$ are the Fourier anti-transforms of the thermal weights $\hbar/i\sinh(\beta \hbar \omega/2)$ and $1/\cosh(\beta \hbar \omega/2)$, respectively:
\begin{align}
	\label{eq:smoothing}
	g_{\Omega}(t)  = \frac {\Omega}{\pi} \frac 1{\cosh(\Omega t) } 
	\ ,\,
	f_{\Omega}(t) = - \frac 1 \beta \tanh(\Omega t  )
	\ ,\,
	\tilde f_{\Omega}(t)  =  \frac  \Omega {\cosh^2(\Omega t )}\ .
\end{align}	
From the first to the second line of Eq.\eqref{eq:invetFDT_R} we have used integration by parts and the definition of $R''_{AB}(t)$.
One can write as well a relation directly connecting the fluctuations and the response, as done by Pottier and Mauer in Ref.\cite{Pottier2001Quantum}.  The inverse transformations and the Pottier and Mauer equations may be checked by applying $\hat {\mathcal L} _{s/c}$ on the $\epsilon$-\emph{regulated} functions $f_\Omega(t+i\epsilon)$, $g_\Omega(t+i\epsilon)$, see the Appendix for all the details.
 We will refer to the equations Eq.\eqref{eq:FDTC}, Eq.\eqref{eq:invetFDT} and Eq.\eqref{eq:smoothing} as the \emph{time-domain FDT}, or $t$-FDT, summarized pictorially in Fig.\ref{fig:tFDT}. 

The functions $g_{\Omega}(t)$ ($\tilde f_{\Omega}(t)$) in Eq.\eqref{eq:smoothing} have a maximum at $t=0$ and then decay exponentially at large $t$ with a width $1 /\Omega$ ($1/2\Omega$), see also Fig.\ref{fig:expPsi} below. 
We call them  \emph{``blurring functions''} because, for finite $\Omega$, they blur out the steepest details of the integrated response and fluctuations.
In fact, in the classical limit $\beta \hbar \to 0$ [$\Omega \to \infty$] they become delta functions,   $\lim_{\Omega \to \infty}g_{\Omega}(t)=\delta(t)$ and $\lim_{\Omega \to \infty}\tilde f_{\Omega}(t) = 2\delta(t)$.  In this limit, the second integral of Eq.\eqref{eq:invetFDT_R} (evaluated at $t'=-t$) vanishes and we remain with
	\begin{align}
	\begin{split}
		F_{AB}(t)  & = C_{AB}(t) \ ,
		 \\
		F_{AB}(t) & = \frac{\Psi_{AB}(\infty)}{\beta}
	- \frac {1}{\beta} \int_{0}^t {\rm d} t'R_{AB}(t')
	\end{split}
	\end{align} 
that corresponds to the classical FDT.

Summarizing, from the $t-$FDT one concludes that quantum effects arise in the time-domain by blurring out, in going to $F_{AB}(t)$,  the fine details of the response and the correlations.

\section{OTOC as two-time functions in a replicated space}
\label{sec:doubledOTOC}

We now discuss how also OTOC obey the quantum FDT Eq.\eqref{eq:fdt} at twice the temperature.
Following Ref.\cite{Maldacena2016bound}, we start with the four-point out of time-order correlator for $t>0$
\begin{align}
	\label{eq:FAB}
	S^{\text{OTOC}}_{AB}(t)=   & \text{Tr} \left ( \rho^{1/2} A(t)B \rho^{1/2} A(t) B\right )  \ ,
\end{align}
with $A(t) = e^{i H t/\hbar} A e^{-i H t/\hbar}$ and $B$ hermitian operators and $\rho = e^{-\beta H}/Z$ a thermal density matrix at inverse temperature $\beta$ and $Z=\text{Tr}e^{-\beta H}$. 

Let us write Eq.\eqref{eq:FAB} in the spectral representation of the Hamiltonian $H$ in terms of  $|i\rangle$ its $i$-th eigenvector  ($H |i\rangle = E_i |i\rangle$ with $|i\rangle \in \mathcal H$, the Hilbert space of the theory)
\begin{align}
	S_{AB}^{\text{OTOC}}(t&) = 
	 \frac 1Z \sum_{ij} e^{-\frac \beta 2(E_i+E_j)} 
		\langle i|A(t) B |j \rangle\, \langle j|A(t) B |i \rangle
	\\
	 = \frac 1Z \sum_{ij}& e^{-\frac \beta 2(E_i+E_j)} 
		\langle i|_1A^1(t) B^1(0) |j \rangle_1\, \langle j|_2A^2(t) B^2(0) |i \rangle_2 \nonumber
	\\
	 = \frac 1Z \sum_{ij}& e^{-\frac \beta 2(E_i+E_j)}
		\langle ij| A^1(t) \otimes A^2(t) \,\, B^1(0) \otimes B^2(0) |ji \rangle \ , \nonumber 
\end{align}
where the states $|ij\rangle = |i\rangle_1 \otimes |j\rangle_2$ live in the double Hilbert space $\mathcal H \otimes \mathcal H$, and are the eigenvectors of the Hamiltonian 
$\mathbb H = H \otimes \mathbb I+ \mathbb I \otimes H$ for the replicated system. 
We also define the operators $\mathbb A(t) = A(t)\otimes A(t)$, $\mathbb B = B \otimes B$ which act in the same replicated space.
By defining a \emph{swap operator} between the two spaces as $\mathbb P |ij\rangle = |ji\rangle $
we can write that expectation as:
\begin{align}
\begin{split}
S_{AB}^{\text{OTOC}}(t) 
    & = \frac{1}{Z} \sum_{ij} e^{- \frac \beta 2(E_i+E_j)} \langle i j |\mathbb{A}(t) \mathbb{B}(0) | j i \rangle
	\label{eq:F1DoubleOperP}
  = \frac 1Z \text{Tr} \left [
	e^{-\beta_2\mathbb H} \, \mathbb A(t) \, \mathbb B\,  \mathbb P 
	\right ] \equiv \mathbb S_{A,BP}(t)   \ .        
\end{split}
\end{align}
This object is, up to a multiplicative constant \footnote{\label{note:constant}The partition function $Z=Z(\beta)=\text{Tr}(e^{-\beta H})$ is not the correct normalization for $e^{-\beta_2 \mathbb H}$, that shall rather be $Z^2(\beta/2)$. Hence Eq.\eqref{eq:F1DoubleOperP} and the ones below differ from equilibrium correlations at $\beta_2$ by the multiplicative factor $Z^2(\beta/2)/Z(\beta)$, a well known constant in the literature, see e.g. Ref.\cite{cotler2017black}.}, the standard structure factor at inverse temperature $\beta_2=\beta/2$, as discussed above in Eq.\eqref{eq:Rsa}. Note that  $\mathbb B\,  \mathbb P$ is also Hermitean, but couples the two spaces $\mathcal H \otimes \mathcal H$.

Inspired by Eq.\eqref{eq:CRdef}-Eq.\eqref{eq:Fsa}, we define the usual  ${\mathbb C}$ and ${\mathbb R}$ as the standard fluctuation and response functions in a double space at inverse temperature $\beta_2$, i.e.
\begin{subequations}
\label{eq:CR}
\begin{align}
	{\mathbb C}_{A, BP}(t)  =&
    \frac 12 \frac 1Z\text{Tr} \left [ {e^{-\beta_2 \mathbb H} }\,\{ \mathbb A(t) , \, \mathbb B\, \mathbb P\}	\right ]  \ ,
    \\
	{\mathbb R}_{A, BP}(t)  =& \frac i \hbar \theta(t) \frac 1Z \text{Tr} \left [ {e^{-\beta_2 \mathbb H} }\, [\mathbb A(t) ,\, \mathbb B\, \mathbb P]	\right ] \ ,
\end{align}    
\end{subequations}
that are related to the real and to the imaginary part of 
${\mathbb S}_{A, BP}(t)= 
	\mathbb C_{A, BP}(t) + \hbar R_{A, BP}''(t)$.
Within this framework, the regularized OTOC in Eq.\eqref{eq:Malda}, for which the bound was proved \cite{Maldacena2016bound}, is:
\begin{align}
    \label{eq:Fgrassa}
    	{\mathbb F}_{A, BP}(t)  =& \frac 1Z \text{Tr} \left [ {e^{-\frac{\beta_2}{2} \mathbb H} }\,\mathbb A(t)    {e^{-\frac{\beta_2}{2} \mathbb H} }     \, \mathbb B\, \mathbb P	\right]
    	\ .    
\end{align}
In this way, the correlation functions ${\mathbb C}(t) ,{\mathbb R}(t) ,{\mathbb F}(t) $ can be pictured as standard two-point fluctuations and responses, in the double space $^{\ref{note:constant}}$. 
As such, they are related through the FDT in Eq.\eqref{eq:fdt} with inverse temperature $\beta_2=\beta/2$ \cite{ueda1}. 

As a remark, we note that the swap operator commutes with the Hamiltonian $[\mathbb H, \mathbb P]=0$, as well as with the operators $\mathbb A(t)$ and $\mathbb B$. Therefore, the operators are  block diagonal in \emph{the symmetric and antisymmetric sectors} $|ij\pm\rangle = (|ij\rangle\pm |ji\rangle)/\sqrt 2$  
and the trace becomes a sum in the two blocks in this basis:  
\begin{align}
\label{eq:plusminus}
{\mathbb S}_{A, BP}(t)& = \frac 1Z \text{Tr} \left [ e^{-\beta_2 \mathbb H} \, \mathbb A(t) \, \mathbb B \mathbb P\,	\right ] ={\mathbb S}_{AB}^+(t) - {\mathbb S}_{AB}^-(t) \ ,
\end{align}
with ${\mathbb S}_{AB}^\pm(t) = \frac 1Z \text{Tr}^\pm \left [ e^{-\beta_2 \mathbb H} \, \mathbb A(t) \, \mathbb B\,	\right ] $ and similarly for all others. 
All FDT relations are hence respected separately by
the ${\mathbb C}^+ ,  {\mathbb R}^+, {\mathbb F}^+$ and the ${\mathbb C}^- ,  {\mathbb R}^-, {\mathbb F}^-$.
Note that, even if $B=A$, correlations in the replicated space are always evaluated for two different operators 
${\mathbb A}$ and ${\mathbb P} {\mathbb A}$ with $\mathbb S_{A, AP}=\mathbb S_{AP,A}$.
On the other hand, in the $\pm$ spaces the correlators involve a single operator $A$, i.e. $\mathbb S^{\pm}_{AA}$. 
One can also check that in the $\pm$ spaces the correlations are just two-time functions plus and minus the four-point functions above, i.e. $\mathbb S^\pm_{AB}(t) = \text{Tr} (e^{-\beta_2 H}A(t) B))^2 / Z \pm \mathbb S_{A, PB}(t)$. 
\\

We now discuss a simple derivation of the bound on $\lambda$ in Eq.\eqref{boundL}.
One can express $\mathbb C(t)$ and $\mathbb F(t)$ 
in the original space as
\begin{subequations}
\label{eq:OTOC_original}
\begin{align}
	\label{eq:OtocSquareLO}
	 \mathbb C_{A, BP}(t) = &
	  \text{Tr} \left ( A(t) \rho^{1/2} A(t) B \rho^{1/2} B  \right ) 
	 - \frac {1}{2} \text{Tr} \left ( \left ( \rho^{1/4} i [A(t), B] \rho^{1/4}\right)^2  \right )\ ,
	 \\
	\label{eq:OtocSquareK}
	{\mathbb F}_{A, BP}(t)
	= & {
	 \text{Tr} \left (\rho^{1/4} A(t) \rho^{1/4} A(t) \,\, \rho^{1/4} B \rho^{1/4} B \right )
	} 
- \frac 12 \text{Tr} \left( \left (  i [\rho^{1/8}A(t)\rho^{1/8}, \, \rho^{1/8}B\rho^{1/8}]\right)^2  \right ) 
\ ,\\
\label{eq:OtocR}
{\mathbb R}_{A, BP}(t)= & i \frac {\theta(t)}\hbar \text{Tr} \left (\rho^{1/2} \{ A(t), B\} \rho^{1/2} \, [ A(t), B ]\right )  \ ,
\end{align}
\end{subequations}
where the right hand side of Eq.\eqref{eq:OtocR} has been discussed in Ref.\cite{tsuji2019out} as a retarded OTOC.
 As already mentioned, the bound on the growth rate in time has been proven in Ref.\cite{Maldacena2016bound} for $\mathbb F(t)$.
  We follow their physical inputs and we assume, firstly, that there is a ``collision time'' $t_d$ after which the two-point functions factorise:
	\begin{align}
	 \text{Tr}& \left ( A(t) \rho^{1/2}  A(t) B \rho^{1/2}  B  \right ) 
	\sim \text{Tr} \left (\rho^{1/2}  A \rho^{1/2}  A\right )  \text{Tr} \left (  \rho^{1/2} B\rho^{1/2}  B  \right ) \ ,
	 \\
	\text{Tr}& \left (\rho^{1/4}  A(t) \rho^{1/4} A(t) \,\, \rho^{1/4} B \rho^{1/4} B \right ) 
	 \sim
	 \text{Tr} \left (\rho^{3/4}  A \rho^{1/4} A  \,\right) \,  \text{Tr} \left (\rho^{3/4} B \rho^{1/4} B \right ) \ .
	\end{align}
This can also be shown using the eigenstate thermalizatation hypothesis ansatz.
 We further assume that there exists an intermediate regime of times $t_d\ll t \ll t_{Ehr}$ (with $t_{Ehr}$ defined below), where the square commutators in Eqs.\eqref{eq:OtocSquareLO}-\eqref{eq:OtocSquareK} grow exponentially in time with the rate $\lambda$, as
 \begin{subequations}
	\begin{align}
		\label{eq:expsc}
		 \text{Tr} \Big(\left (\rho^{1/4} i[A(t), B]\rho^{1/4} \right )^2\Big ) & = 2 D \delta e^{\lambda t}\ ,  \\
		 \text{Tr} \left ( \left (  i [\rho^{1/8} A(t) \rho^{1/8}, \, \rho^{1/8}B\rho^{1/8}]\right)^2  \right )
		 &= 2 D \epsilon e^{\lambda t} \ ,
	\end{align}
 \end{subequations}
	with $\epsilon$ and $\delta$ two positive and small constants, and $D$ a positive order one constant that depends on the specific operators.   
Within our notations, via Eqs.\eqref{eq:OtocSquareLO}-\eqref{eq:OtocSquareK}, these approximations for $B=A$ read
\begin{subequations}
\begin{align}
\label{Creg}
	 \mathbb C_{AA}(t) & = \left ( \text{Tr}(\rho^{1/2} A \rho^{1/2}A)\right )^2 -  D\delta e^{\lambda t} \ ,\\
\label{Freg}
	 \mathbb F_{AA}(t) 
	&  = \left ( \text{Tr}(\rho^{1/4} A\rho^{3/4}A)\right )^2 -  D e^{\lambda (t-t_{Ehr})} \ ,
\end{align}	
\end{subequations}
where the Ehrenfest time is
$
	t_{Ehr} = \frac 1 \lambda \log \epsilon^{-1}
$. Everything is well-defined for systems  with $\epsilon \ll 1$ (like semi-classical $\epsilon =\hbar$ or large $N$ models with $\epsilon=1/N^2$).	\\

 A short computation  with (\ref{eq:FDTC}) shows that if ${\mathbb F}(t)\sim e^{ \lambda t}$ then the $t$-FDT immediately implies \cite{Foini2012Dynamic}:
\begin{align}
\begin{split}
{\mathbb C}(t)& = \cos \left (\frac {\beta_2 \hbar}2  \lambda \right ) {\mathbb F}(t) \ ,\,
 {\mathbb R}(t)  = - \frac{ 2}{\hbar} \theta(t) \sin \left (\frac {\beta_2 \hbar}2  \lambda \right ) {\mathbb F}(t) \ .
\label{laura}
\end{split}
\end{align}
(This is in fact true for any $R,F,C$ with exponential time-dependence, not necessarily derived from a space-doubling).
As observed by Tsuji et al. \cite{ueda1,ueda2} in a slightly different formulation, if in Eq. (\ref{laura}) the sign of the cosine is negative, this leads to a contradiction with (\ref{Creg}) and (\ref{Freg}), since both exponential terms are by their definition positive definite.
 We thus conclude that the Lyapunov exponent $\lambda$ of $\mathbb F$ must be such that $\cos\left(\frac{\beta_2 \hbar}{2} \lambda\right)\geq 0$, leading to the bound in Eq.\eqref{boundL}. 
 In all Planckian models (including SYK \cite{maldacena2016remarks}) the quantity $\frac{\hbar\lambda}{T}$
grows as $T$ decreases, so it is always included in the first quadrant with positive cosine, hitting its upper boundary, if the bound is reached, at $T=0$ \cite{Kurchan2018Quantum, pappalardi2021low}.

As we have recognized here, the FDT relations discussed by Tsuji et al. in Ref.\cite{ueda1, ueda2}, that provide the bound on chaos, can be understood as simple FDT/KMS relations for two-point functions. Notably, the factor $\cos\left(\frac{\beta_2 \hbar}{2} \lambda\right)$ of Eq.\eqref{laura} (discussed at length below) also appears in the relation between the magnitude and the exponent of OTOCs derived by Gu and Kitaev in a diagrammatic approach in Ref.\cite{gu2019relation}.
We are thus encouraged to try to understand the underlying mechanisms by which the quantum FDT imposes bounds on exponentially decreasing or increasing two-point correlations.
In what follows, we shall obtain a `constructive' explanation of the impossibility of the negative proportionality constant between ${\mathbb C}$ and ${\mathbb F}$.


\section{Consequences of the KMS conditions on two-time functions}
\label{sec:sei}
\label{sec:sei}
In this section, we will study the effects of the quantum FDT on two classes of correlations and responses, namely those that depend exponentially on time with a positive or negative rate.
For simplicity we shall concentrate on two-point functions associated with one single operator $\langle A(t) A\rangle$ or $\langle \mathbb A(t) \mathbb P \mathbb A\rangle$ (and neglect the sub-indices, i.e. $C(t)=C_{AA}(t)$ etc.).

\subsection{Realizable correlations and responses} 
\label{Sec_realizable}
One is first led to the question as to what are the possible physically realizable correlation
and response functions and whether we can hope to derive a universal bound for their rate of decay or growth.
\begin{itemize}
    \item In the frequency domain, $\omega \text{Im}R(\omega)$ - the average work done on a system by an oscillating field of frequency $\omega$, has to be positive for all $\omega$
    to satisfy the Second Principle. Integrability in $\omega$ and zero frequency limits are imposed by the finiteness of moments and of the static susceptibility \cite{forster}.
   Beyond this,  \emph{any} response with $\text{Im}R(\omega)$ which satisfies these conditions  is physically realizable.    The same is true about $C(\omega)$. 
   An explicit example is a set of infinitely many oscillators (for $\text{Im} R(\omega)$) or fermions (for $C(\omega)$), with an appropriate distribution of characteristic frequencies.
\end{itemize}    
To see this, consider a set of free bosons $ H=\sum_i \hbar \omega_i a^{\dag}_i a_i $
with distribution of characteristic frequencies $P(\omega)$, obviously positive definite. Choosing ${R(t,t')= \frac 1 N \sum \text{Tr} \; ( e^{-\beta \hat H}[\hat x_i(t), \hat x_i(t')] )\theta(t-t')}$ with $x_i = \sqrt{\frac{\hbar}{2\omega_i}} (a^\dag_i + a_i)$ one has $P(\omega) = \omega {\mbox{Im}} R(\omega)$ (for $\omega>0$).  This implies that any response function, and therefore via the FDT any correlation  $C(\omega)$, can be realized with a suitable chosen set of harmonic oscillators. \footnote{A similar argument may be done considering a free fermionic system, with the $a_i$ being now fermions: defining the retarded function, one in fact concludes that the Fourier transform of the anticommutator is directly related to the distribution $P(\omega)$ used to generate the desired functions.}
\begin{itemize}
    \item  As we show below, bounds on the exponential growth or decay of $R(t)$ and $C(t)$ apply within certain conditions. Hence, the previous remark implies the non-existence of universal bounds without \emph{further assumptions}. 
    \item  The example of ${\mathbb R}$ and ${\mathbb C}$ in Eq.\eqref{eq:OTOC_original} --- which have a particular form in a replicated space [cf. Eq.\eqref{eq:OTOC_original}] \footnote{In this case, the second principle leads to the condition for $\omega \text{Im} R_{A, PA}(\omega)$ being a positive definite form \cite{Fabrizio}. This implies the positivity of the $\omega \text{Im} R^{\pm}_{AA}(\omega)$ in the (anti)symmetrized spaces [cf. Eq.\eqref{eq:plusminus}].}--- is an instance of a further condition (the same sign between $\mathbb F$ and $\mathbb C$ in the relevant  time-domain) that implies a universal bound on the growth rate.
\end{itemize}

\subsection{Blurring}
\begin{figure}[t]
\begin{center}
\includegraphics[width=7.2cm]{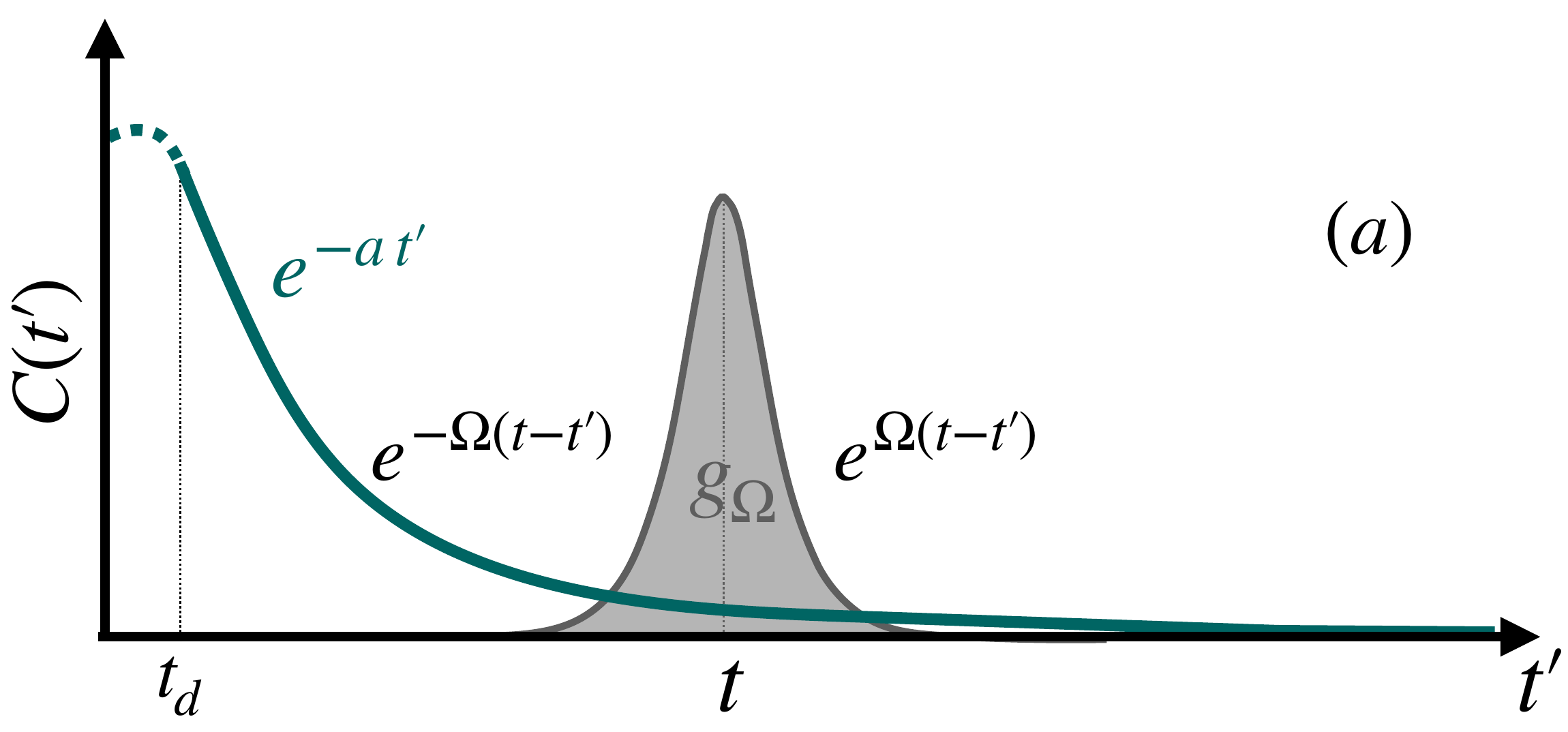}
\includegraphics[width=7.2cm]{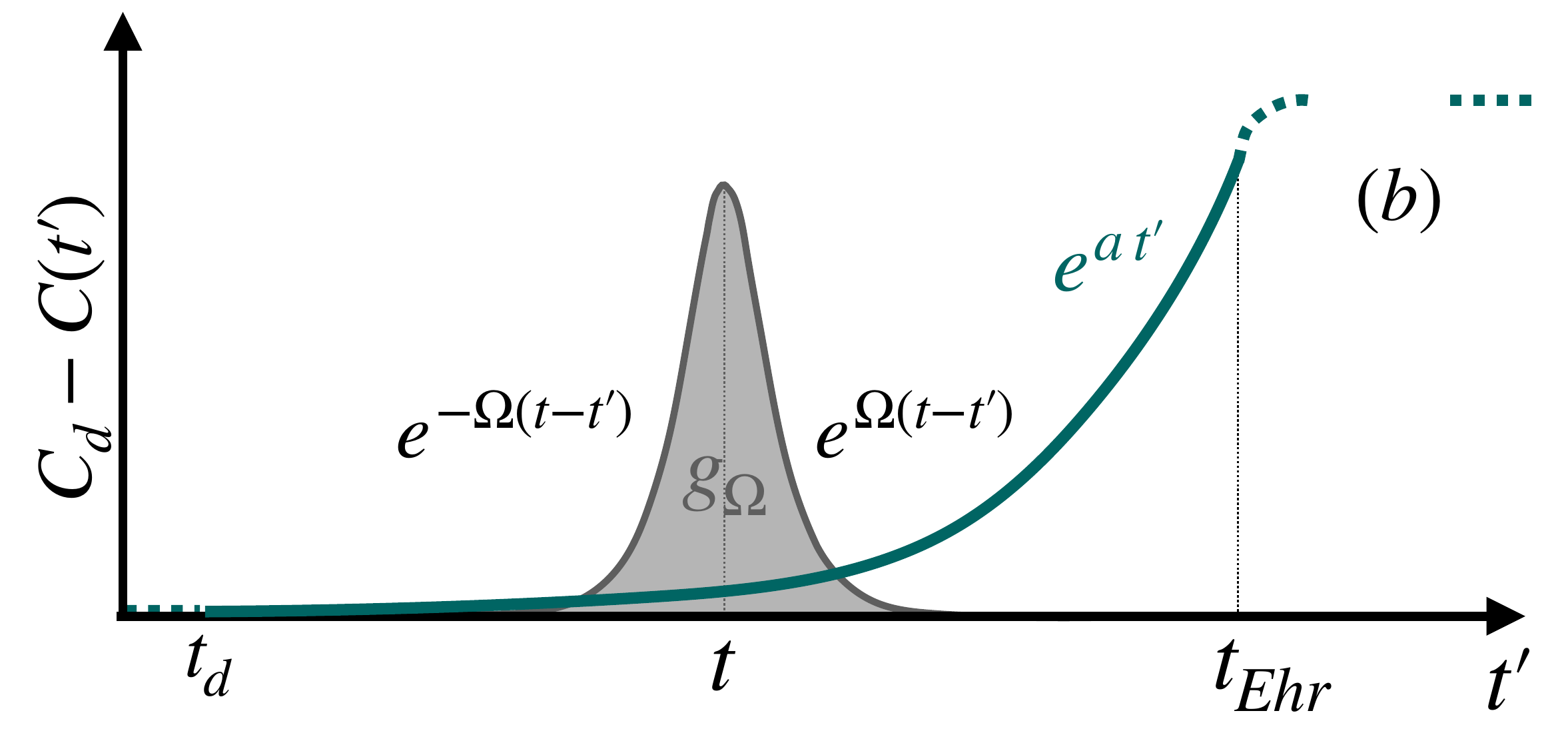}
\caption{$t$-FDT on fluctuations $C(t)$ that depend exponentially on time. (a) A fluctuation function $C(t')$ that decays exponentially with a rate $a$. The  dashed line indicates the early time (model-dependent) regime. (b) A fluctuation increases exponentially in time for $t_d\ll t\ll t_{Ehr}$ with rate $a$. What happens in the post-Ehrenfest times is, again, model-dependent.
The competition between 
the exponential decay of $\cosh\left({\Omega t}\right)$ and the exponential dependence of $C$ decides which time-interval dominates the integral. When the exponents in $C$ become large enough compared to $\Omega$,
there is a transition from a small `blurring', to a situation where either early or post-Ehrenfest times
dominate.
A similar picture holds for the response function $R(t)$. We thus clearly see from the picture that the result of the convolution can  be negative only if it is the boundary times that dominate.}
\label{fig:expPsi}
\end{center}
\end{figure}

As discussed  above, the quantum $t-$FDT in Eq.\eqref{eq:invetFDT}-(\ref{eq:smoothing}) acts on $F(t)$ as a blurring of the fine details of the time-dependent correlations on a timescale $\tau_{\Omega} = \Omega^{-1}$ Eq.\eqref{Omega} that shrinks to zero in the classical limit/high temperature regime. 
In what follows, we study how this affects correlation functions that depend exponentially on time, see e.g. Fig.\ref{fig:expPsi}.
In particular, we shall consider two ansatz of the form:
\begin{equation}
    \label{eq:ans1}
    C(t) = D e^{- a t} 
    \qquad \text{or}\qquad
    R(t) = D e^{- a t}
    \qquad \text{for}\qquad t\gg t_d
\end{equation}
where $D$ is a constant and $t_d$ is a miscoscopic timescale setting the crossover to the exponential regime, before which we do not make any assumption. We also consider
\begin{equation}
\label{eq:ans2}
  C(t) = C_d - D \epsilon e^{a t}
   \quad\text{or}\quad
  R(t) = D \epsilon e^{a t}
  \quad\text{for} \quad t_d\ll t \ll t_{Ehr}
\end{equation}
with $t_d$ and $t_{Ehr}$  two parametrically distant times which allow for the definition of an intermediate exponential regime and $\epsilon$ a small parameter. Sometimes it is easier to express the $t$-FDT in terms of the integrated response $\Psi(t)$ in Eq. (\ref{Eq_psi}), which inherits the exponential dependence on time of $R(t)$.

While we give generic arguments for two-point functions with an exponential dependence on time, the ansatz (\ref{eq:ans2}) can describe as well the two-times functions $\mathbb C$, $\mathbb R$ and $\mathbb F$, that we have defined above for OTOC, with now $a=\lambda$ the Lyapunov exponent. They are, however, a particular case coming from replication of Hilbert space, which, in addition, has an inverse temperature  $\beta_2=\beta/2$ -- $\beta$ being the temperature entering in the definition of the associated OTOC functions --  and the blurring of the $t$-FDT occurs on a scale $2\tau_\Omega$. 

In the case of exponential decaying correlation whose Fourier transform is known, one can perhaps study the problem exclusively in the frequency domain, but as we shall see, for growing exponential this becomes
less clear and time-domain calculations are preferred. In the Appendix we consider two toy models with a Lorentzian fluctuation $C(\omega)$ [and response $\text{Im}R(\omega)$] functions, that are often described in the literature as phenomenological models for transport \cite{Allen2006Conceptual, forster}.

 The effect of blurring can be summarized as follows. Starting from the generic exponential ansatz for $C(t)$ Eq.\eqref{eq:ans1}-Eq.\eqref{eq:ans2}, we will show how the $t$-FDT results in a bound on the rate of $F(t)$. Via the differential operator $\hat{\cal{L}}_{c}$, the bound will then be inherited by the conjugate quantity  $R(t)$; and, conversely, starting from a generic $R(t)$, to $C(t)$ via $\hat {\cal{L}}_{s}$.
These bounds are established under some conditions, whose genericity and stability we shall discuss.

\subsection{Exponential decay}
We start by considering \emph{fluctuations} that decay exponentially at large times as
\begin{equation}
        C(t) = D e^{-a t} \quad \text{for} \quad t\gg t_d \ ,
\end{equation}    
with some rate $a >0$. 
One can see the bounds emerging by evaluating the most relevant contributions to the integral in Eq.\eqref{eq:invetFDT_C}.
For $a < \Omega$, the blurring function $ g_\Omega(t-t')$  is sharply peaked with respect to the variation of $C(t')$, and thus the integral is dominated by times $t'\sim t$.
For $a > \Omega$, there is a transition to a situation in which the integral is dominated by early times (the blurring is strong), a regime in which the exponential approximation of $C$ does not hold, but where one is allowed to expand the  $g_\Omega(t-t') \propto 1/\cosh\Omega(t-t')$ in exponentials $e^{-\Omega (t-t')}$. 
All the details of the computation on the integral of Eq.\eqref{eq:invetFDT_C} can be found in the Appendix. The leading contributions to $F(t)$ at large times are 
\begin{align}
\label{eq:Ft}
    F(t) & \simeq  \frac{D}{\cos(a \pi / 2\Omega)}  e^{-a t} +  c_{\Omega} e^{- \Omega t} + \mathcal O(e^{-3 \Omega t})  \sim e^{-t /\tau} \ ,
\end{align}
where $\tau$ is defined as the decay rate of the dominant term for $F$.
Here, for $a>\Omega$ the coefficient $c_\Omega$ is given by
\begin{equation}
\label{eq:cn_def}
 c_{(1+2n) \Omega} = \frac{2\Omega}{ \pi}  \int_{-\infty}^{\infty}{\rm d} t'\,  C(t') e^{(1+2n) \Omega t'} \ ,
\end{equation}
with $n=0$, while for $a<\Omega$  the $c_{\Omega}$ is given in the Appendix. Notice that the coefficient in front of $e^{-at}$ in Eq.\eqref{eq:Ft} is the one we would have obtained by naive inversion of Eq.\eqref{laura}. By applying the differential $\hat {\mathcal L}_s$ on Eq.\eqref{eq:Ft}, one has:
\begin{align}
    \label{eq:Ft1}
    R(t) &  \simeq 
    \frac 2 \hbar \left (   {D} \tan \left (\frac{a \pi}{2 \Omega} \right )  {e^{-a t}} + c_\Omega e^{- \Omega  t}   + \dots
     \right ) \sim e^{- t/\tau} \ .
\end{align}
Hence, if $c_{\Omega} \neq 0$, the rate of exponential decay of $F(t)$ and $R(t)$ is bounded by
\begin{equation}
    \label{eq:doubleBound}
    \frac 1 \tau \leq \Omega = \frac{\pi}{\beta \hbar} \ .
\end{equation}
Notice that the nature of the fluctuations at early times (the ultra-violet behaviour here supposed unknown) is essential to evaluate the long-time behaviour of $F(t)$, since it determines the $c_\Omega$ in Eq.\eqref{eq:cn_def}. 
The assumption  $c_{\Omega} \neq 0$ is central, and we shall come back to this point in the discussion below.

Suppose, instead, that $c_\Omega=0$. In this case the rate of $F$ and $R$ have no bound at $\Omega$. One could imagine to start from a \emph{response function} that decays exponentially at large times, i.e. $R(t)= D e^{-a t}$ for $t\gg t_d$ with a generic rate $a >0$.  
The resulting integrated response goes as
$    \Psi(\infty)-\Psi(t) = D\frac {e^{-a t}}a$
for $t\gg t_d$.
One now determines the $F(t)$ from the integral in Eq.\eqref{eq:invetFDT_R} and the response via the differential $\hat {\mathcal L}_c$. Exactly the same reasoning outlined above applies, where now the smoothing function $\tilde f_\Omega(t)$ has a width $1/2\Omega$, see the Appendix. The blurring of the $t$-FDT and the differential operator lead to

\begin{subequations}
\label{eq_F}
\begin{align}
    F(t) \simeq
    \frac \hbar 2 \frac{D}{\sin(a \pi / 2\Omega)}  \,{e^{-a t}}
    + \, r_{2\Omega} \, e^{- 2\Omega t} + \dots 
     \sim e^{- t/\tau}\ ,
    \\
    C(t) \simeq  
   \frac \hbar 2 D \, \cot \left ( \frac{2a}{\pi \Omega}\right ) \, e^{-a t}
    -  r_{2 \Omega} e^{-2\Omega t} + \dots
    \sim e^{- t /\tau} \ .
\end{align}
\end{subequations}
where for $a>2\Omega$ the coefficient $r_{2\Omega}$ is given by
\begin{equation}
\label{eq:r2n_def}
r_{2n\Omega} = \frac 4 \beta \int_{-\infty}^\infty {\rm d}t'  \, iR''(t)\,  e^{2n\Omega t} 
\end{equation}
for $n=1$ and $R''(t)=\langle [A(t), A]\rangle/2\hbar$ related to $R(t)$ via Eq.\eqref{eq:Csa}. 
 For $a<2\Omega$  the $r_{2\Omega}$ is given in the Appendix. From this expression we find that for $r_{2\Omega} \neq 0$ the rate of the exponential decay of the $F(t)$ \emph{and} of the fluctuations are bounded by
\begin{align}
    \label{bound}
    \frac 1 \tau \leq  2\Omega  = \frac {2\pi}{\beta \hbar} \ .
\end{align}
Also in this case, the assumption $r_{2\Omega} \neq 0$ is crucial and it is discussed below.

\subsection{Exponential growth}

The same blurring effect induces bounds for the exponents of correlations functions that increase exponentially in a time interval. We may start from a correlation function that grows exponentially in an interval:
\[
C(t) = C_d -D\epsilon e^{at} \quad \text{for} \quad t_d\ll t \ll t_{Ehr} \ ,
\]
as illustrated in Fig.\ref{fig:expPsi}b. We consider a small parameter $\epsilon = e^{-a t_{Ehr}}\ll 1$, i.e. the Ehrenfest time is large enough $t_{Ehr}\gg t_d$. Exactly as done for the decaying correlation functions, bounds can emerge by evaluating the most relevant contributions to the integral in Eq.\eqref{eq:invetFDT_R} in the interval $t_d\ll t \ll t_{Ehr}$. 
There are two possibilities mirroring the ones discussed before: if $a<\Omega $ the integral is dominated by times $t\sim t'$, while with $a>\Omega$ there is a transition to a regime
in which  the blurring is strong: the integrand grows throughout the region $t_d \ll t \ll t_{Ehr}$ and is dominated by times $\gtrsim t_{Ehr}$, where the behavior of $C(t')$ is not specified. The evaluation leads to (see Appendix):
\begin{subequations}
\begin{align}
     & C_d - F(t) = D \frac{e^{a (t-t_{Ehr})}}{\cos(a \pi / 2 \Omega)} + C_\Omega e^{ \Omega  (t-t_{Ehr})} + \dots  
     \simeq e^{\lambda(t-t_{Ehr})} \ , 
\end{align}
\end{subequations}
where $\lambda$ is defined as the dominant rate to $F(t)$. For $a>\Omega$ the coefficient is given by
\begin{equation}
    C_{(1+2n)\Omega}  =  \frac{2 \Omega}{\pi} \int_{t_d}^{\infty}  \left (C_d - C(t') \right) e^{-(1+2n)\Omega(t'-t_{Ehr})} {\rm d} t'\ .
\end{equation}
with $n=0$, while for $a>\Omega$ we refer to the Appendix.
If we  assume that {\em every $C_{\Omega}$ is of order one}, then within the interval $t_d\ll t\ll t_{Ehr}$ the lowest of $(\Omega,a)$ dominates in  the Lyapunov regime (\ref{asymptotic}).  
The assumption of $C_{\Omega}=O(1)$ is based on the fact that  $C_d-C(t')$ grows exponentially up to $t_{Ehr}$ and it eventually saturates. Hence, it is reasonable to assume that the integral is dominated by times around $t_{Ehr}$ and that $C_\Omega=\mathcal O(1)$. However, at this level, this remains in general only an assumption. Hence, provided $C_\Omega\neq 0$, applying $\mathcal L_s$ one obtains that the exponential rate in $F(t)$ and $R(t)$ is bounded by
\begin{align}
    \label{ok}
    \lambda \leq \Omega  = \frac {\pi}{\beta \hbar} \ .
\end{align}
Note that for $C=\mathbb C$ and $F=\mathbb F$, we have $a=\lambda$ and this equation shall be evaluated at $\beta_2=\beta/2$, leading to the correct bound on the quantum Lyapunov exponent.

If instead $C_\Omega = 0$, one may consider a response function that grows exponentially as $R(t) = D \epsilon e^{a t}$ in an interval $t_d\ll t \ll t_{Ehr}$, after which its behaviour is unknown.
  The associated integrated response is
$
    \Psi(t) \simeq \Psi(t_d) + \frac Da e^{a(t-t_{Ehr})} 
    \quad \text{for} \quad t_d\ll t \ll t_{Ehr} \ ,
$
and for larger times it eventually saturates to its static susceptibility $\Psi(\infty)$. 
We study the effect of the convolution with $\tilde f_\Omega$ on $\Psi(t')$.  Repeating the same arguments (see Appendix):
\begin{align}
    \label{asymptotic}  
    & \frac {\Psi(\infty)-\Psi(t_d)}\beta  -  F(t)  
    \simeq e^{\lambda(t-t_{Ehr})} 
    \\ 
   & \quad \simeq
     \frac \hbar 2 D \frac{e^{a (t-t_{Ehr})}}{ \sin(a \pi /2\Omega)} 
     + R_{2 \Omega} e^{2\Omega (t-t_{Ehr})} + \mathcal O(e^{4 \Omega (t-t_{Ehr})})\nonumber
   \ ,
\end{align}
being $\lambda$ the rate of the dominant term in $F(t)$. The coefficient for $a>2\Omega$ is given by
\begin{align}
\begin{split}
    \label{ckays}
     R_{2 n \Omega}
 = \frac 1 \beta 
 \int_{t_d}^\infty R(t')
        e^{-2n \Omega(t'-t_{Ehr})}\;\; 
 \ .               
\end{split}
\end{align}  
with $n=1$.
Also in this case, it is reasonable to assume that the integral is dominated by times around $t_{Ehr}$ and that $R_{2\Omega}=\mathcal O(1)$. At this level, this remains an assumption. 
If it holds, applying $\hat {\cal{L}}_c$ we conclude that the rate of exponential growth of $F(t)$ and $C(t)$ is bounded by
\begin{equation}
    \label{double}
    \lambda \leq 2\Omega = \frac{2 \pi}{\beta \hbar} \ .
\end{equation}
When evaluated for the OTOC $\mathbb F(t)$ in the double space at $\beta_2 = \beta /2$, the rate $a=\lambda$ is the Lyapunov exponent and Eq.\eqref{double} is twice the usual bound to $\lambda$.

\subsection{The assumptions for the bounds: genericity}

\label{sec:gene}
For the case of exponentially decreasing functions, we have found above various possibilities:
    \begin{enumerate}
    \item \label{case1} Decay rate of $F,\,  R\,$ bounded at  $\Omega$ if
    \begin{align}
        \label{c2n}
        \begin{split}
        c_{\Omega} = \frac{2\Omega}{ \pi}  \int_{-\infty}^{\infty}{\rm d} t'\,  C(t') e^{ \Omega t'}\neq 0 \ ,
        \end{split}
    \end{align}
    i.e.  $C(\omega)$ does not have a zero in Fourier space at $i \Omega$, implying a pole for $F(\omega)$ at the same frequency;
    \item \label{case2} Decay rate of $F,\,  C\,$ bounded at $2 \Omega$ if
    \begin{equation}
        \label{r2n}
        c_\Omega = 0 \ ,\quad
        r_{2\Omega}  = \frac 4 \beta \int_{-\infty}^\infty {\rm d}t'  \, iR''(t)\,  e^{2\Omega t} \neq 0 \ ,
    \end{equation}
     i.e. $ \text{Im}R(\omega)$ does not have a zero in Fourier space in $\omega= 2i \Omega$, implying a pole for $F(\omega)$ at the same frequency;
    \item \label{case3} No arguments for bounds on the decay rates if
    \begin{equation}
        \label{00}
        c_{(1+2n)\Omega} = r_{2n\Omega} = 0 \ ,
    \end{equation}
     i.e. $F(\omega)$ has no poles in either $\omega=2ni\Omega $ and $\omega= (2n+1)i \Omega$. 
\end{enumerate}

Hence, the bounds (or their absence) rely on the zeros of $C(\omega)$ and $R(\omega)$ in the complex plane at the Matsubara frequencies. 
Therefore, one shall question what is the fate of these zeros for generic physical systems. 
A possible way to tackle the issue is to think in terms of the stability of these zeros under small perturbations, which fits a scenario of \emph{stochastic stability} at the level of $C$ or $R$.  The argument arises from applying  {\em generic} classes of perturbations, which will make the integrals in Eq.\eqref{c2n} and in Eq.\eqref{r2n} to become non-zero if they vanished originally. 
This case, where one shall have a $C$ or $R$ as general as possible, implies that the function $F$ has poles in all the Matsubara frequencies, due to the structure of FDT in Eq.\eqref{eq:fdt}. For example, these perturbations may be enforced physically by weakly coupling the system to a bath of oscillators, as in Schwinger-Keldysh. A simple but tedious computation shows that if
the bath itself does not have a zero in response or correlation on a certain complex frequency, but the system has, the perturbation shifts the zero away from that place. The classification of possible baths and their impact on the bounds is an interesting topic for further research. 
 This kind of argument is thus like the one invoked to rationalize level-repulsion for generic chaotic systems, a matter of stochastic stability.
 These have clearly their limitations: for example, an integrable system cannot be perturbed generically and stay integrable.
 Similarly, a system like the ${\mathbb C}$, ${\mathbb R}$, ${\mathbb F}$ deriving from treating an OTOC in replicated space, if modified by
 a generic perturbation in replicated space no longer derives from an OTOC.

If possibility \ref{case3} is stable under perturbation, then the $F$ is the most generic function, without a specific structure at the Matsubara frequencies. Note that in such a case, the vanishing of equations (\ref{eq:cn_def}) and (\ref{eq:r2n_def}) corresponds to sum rules that have to be satisfied by correlation and response.

In the Appendix we discuss some toy models which are in class \ref{case1} or \ref{case2} described above. We did not find simple examples realizing situation \ref{case3} with unbouned rate for $F$, $C$ and $R$, with the constraint of a $F(\omega)$ falling fast at large frequencies \cite{Srednicki1999}.
Nevertheless, at this level, we are not able to rule out that $F$ is generically unbounded in case \ref{case3}.

Let us now turn to the assumptions for exponential growth in an interval of time. 
With similar arguments, we have identified the following conditions on integral quantities:
    \begin{enumerate}
    \item  Growth rate of $F$ and $R$ bounded at $\Omega$ if
    \begin{align}
       C_{\Omega} & =  \frac{2 \Omega}{\pi} \int_{t_d}^{\infty}  \left (C_d - C(t') \right) e^{-\Omega(t'-t_{Ehr})} {\rm d} t'
       \sim  O(1) \ .
 \label{c2n_}
    \end{align}
    \item  Growth rate of $F$ and $C$ bounded at $2\Omega$ if
    \begin{equation}
        \label{r2n_}
        C_\Omega = 0 \ ,\quad 
        R_{2\Omega} =\frac 1 \beta  \int_{t_{d}}^\infty R(t')\, e^{-2 \Omega(t'-t_{Ehr})}   {\rm d} t'
        \sim O(1) \ ,
    \end{equation} 
\end{enumerate}  
Here, by $O(1)$ we mean that they do not vanish parametrically with, e.g. $t_{Ehr}$.  
The main difference with the previous condition is that, since we are looking at time interval $t\ll t_{Ehr}$, the crossover time $t_{cr}$ above which the rate $\Omega$ becomes dominant (defined e.g. by $e^{(a-\Omega)(t_{cr}-t_{Ehr})} = \frac{C_{\Omega}}{D}$) 
might hit the boundary of the interval. For this reason, these conditions are certainly stronger than the ones for a (single) exponential decay, where one may always look at infinite times.

Here we do not have the simple picture of zeroes in Fourier space, but still one could argue for arguments of stochastic stability.
We shall not delve into this matter further, and just take this as a different working hypothesis.

\subsection{Discussion}
\label{Sec_discussion}
Let us \emph{summarize} our set of arguments on the blurring. 
The $t$-FDT leads us to show 
that from a fluctuation $C(t)$(or response function $R(t)$) which depends exponentially on time with unbounded rate and satisfies Eq.\eqref{c2n} (or Eq.\eqref{r2n}) one obtains an intermediate function $F(t)$ which also has an exponential behavior whose rate however is bounded to $\Omega$ (or $2\Omega$) respectively. The same decay and bound is inherited by the conjugate variable $R(t)$ (or $C(t)$) via the differential $t$-FDT.
Summarizing:
\begin{enumerate}
    \item if $C(t)$ unbounded exp. [with $c_{\Omega}\sim C_{\Omega}\sim \mathcal O(1)]$  
    $\longrightarrow F(t)\sim R(t)\sim e^{t/\tau} $ with $|\frac 1\tau|\leq \Omega$ \ ;
    \item if $R(t)$ unbounded exp. [with $r_{2\Omega}\sim R_{2\Omega}\sim \mathcal O(1)]$  
    $\longrightarrow F(t)\sim C(t)\sim e^{t/\tau} $ with $|\frac 1\tau|\leq 2 \Omega$.
\end{enumerate}

These bounds 
depend on the existence of the coefficients $r/R_{2\Omega}$ and $c/C_{\Omega}$. Their genericity has been rephrased in terms of the stochastic stability of responses and fluctuations. On the other hand, when two-points functions satisfy further constraints, like coming from a replicated space as for OTOC, the $t$-FDT can lead to universal bounds, as the one for the Lyapunov exponent Eq.\eqref{boundL}.
\\
 Let us also comment that if $c_\Omega \neq 0$, one can certainly conclude that $F(t)$ and $R(t)$ have a bound at $\Omega$, but nothing can be said about $C(t)$ because we cannot repeat the reasoning from a generic $R(t)$, being the latter bounded.
From  our calculations, it may also happen that $C(t)$ and $F(t)$ have a different exponential decay or growth if $F$ is characterized by a rate $\Omega$. Even if we can not rule out such a situation in general, it would correspond to an intrinsically quantum effect, since in the classical limit $C(t)\sim F(t)$ and the exponential decay or growth of $C(t)$ and $R(t)$ is necessarily the same. 

The only situation which would lead to an \emph{unbounded decay} for $F(t)$, and therefore a decay unbounded both for $C(t)$ and $R(t)$, is such that all the coefficients discussed above $c_{\Omega}$ and $r_{2\Omega}$ vanish [cf. case \ref{case3} in the previous section]. However, it is not clear if one would be able to construct a function $F(t)$ without a bound, under the necessary condition that $F(\omega)$ decays fast enough at large frequencies \cite{Murthy2019Bounds}. Understanding the relationship between the bounds imposed by the structure in frequency \cite{Murthy2019Bounds} and the blurring of the $t$-FDT may lead to a universal bound for $F(t)$ beyond any assumption of stochastic stability.\\

We have explored the role of \emph{FDT in the time-domain}, and how it leads to a smoothing of the rates of change of correlation functions. 
Working with correlation functions over time, rather than in frequency, has significant advantages.
Typically, the hypotheses on dynamical quantities are given in the time-domain, as in the case of Lyapunov regimes. The same is true of systems that have decay regimes identified by different time scales, parametrically separated. In these cases, windows in time become mixed up when translated to Fourier space, and the problem may soon become intractable. 
Another advantage of the $t$-FDT is that growing exponentials in time lead, in frequency, not to true poles (as decreasing exponentials would) but to pseudo-poles which are smoothed by a quantity inversely proportional to the (unavoidable) cutoff time of the exponential regime. For all these reasons, the analysis over time can be preferable. 

 Furthermore, one may interpret the FDT blurring as an \emph{information loss} on the correlation functions. Below the threshold, $\Omega$, from a measurement of $R(t)$ one can always infer the large time behavior of $C(t)$ by naive inversion of (\ref{laura}). On the other hand, the knowledge of the tail (large times) of $R(t)$ does not allow for the reconstruction of the tail of $C(t)$, when the response decay rate is at $\Omega$ (strong blurring regime).

\section{Conclusions}

In this paper, we discussed the quantum FDT as the principle physical underlying the quantum bounds on transport coefficients and Lyapunov exponents, usually saturated by black-hole models. 
Our work started from the observation, related to the work of Tsuji et al. \cite{ueda1, ueda2},  that the bound $\lambda \geq \frac{2 \pi }{\beta \hbar}$ can be seen as inconsistency of the relative sign of two exponentially increasing functions encoded in the structure of OTOC. This follows from the $t$-FDT that relates the OTOC functions ${\mathbb C}(t)$ and ${\mathbb F}(t)$, that in turn, we mapped to two-point functions in a replicated space.
This result motivated us to study the effect of the quantum FDT on generic two-point functions with an exponential (increasing or decreasing) dependence on time.

We have noted that the effect of the FDT is particularly transparent in the time domain because starting from a correlation or (integrated) response it induces a smoothing
of the intermediate regularized function $F(t)$ which is then inherited by the conjugate function.
This blurring acts on a Planckian time scale $\tau_{\Omega}$ that emerges by Fourier transforming to the time domain the hyperbolic functions that characterize the FDT in frequency, Eq.(\ref{eq:fdt}).

Our work paves the way for a unified understanding of the bounds enforced by a quantum mechanism on different physical quantities -- like appropriately normalized transport coefficients --
on the basis of the constraints and the timescale $\tau_{\Omega}$ imposed by the \emph{quantum} FDT.

\section*{Acknowledgements}
We wish to thank M. Fabrizio, J. Maldacena and C. Murthy for useful discussions.  SP and JK are supported by the Simons Foundation Grant. No 454943. This work is supported by ``Investissements d'Avenir" LabEx PALM
(ANR-10-LABX-0039-PALM) (EquiDystant project, L. Foini).

\appendix

\section{Derivation of the $t$-FDT}
\label{app:FDT_time}

In this section, we consider the time-domain version of the standard FDT relations at inverse temperature $\beta$
\begin{subequations}
\label{eq:fdt_a}
\begin{align}
\label{eq:fdtC_a}
	C_{AB}(\omega) & = \cosh(\beta \hbar \omega /2)  F_{AB}(\omega) \ ,\\
\label{eq:fdtR_a}	
	\hbar R''_{AB}(\omega) & = \sinh(\beta \hbar \omega /2)  F_{AB}(\omega) \ ,
\end{align}	
\end{subequations}
where $C(\omega),\, R''(\omega)$ and $F(\omega)$ are the Fourier transform of 
\begin{align}
    C_{AB}(t) = \frac 12 \text{Tr} 
    \left ( \rho \{ A(t), B \} \right ) \ ,
    \quad
     R''_{AB}(t) = \frac 1{2\hbar} \text{Tr} 
    \left ( \rho [A(t), B ] \right ) \ ,
    \quad 
    F_{AB}(t) = \text{Tr} \left (
    \rho^{1/2} A(t) \rho^{1/2} B \right ) \ ,
\end{align}
with $\rho = e^{-\beta H}/Z$ the thermal density matrix.
For the sake of notations, let us fix here the signs of the Fourier transform (FT) and its inverse (aFT)
\begin{align}
	f(\omega) = \int_{-\infty} ^\infty dt e^{i\omega t} f(t) \ , 
	\quad \quad
	f(t) = \int_{-\infty} ^\infty \frac{d\omega}{2\pi} e^{-i\omega t} f(\omega) \ ,
\end{align}
and introduce 
\begin{align}
	\Omega \equiv \frac{\pi}{\hbar \beta} \ ,
\end{align}
that diverges in the naive classical limit $\hbar \beta \to 0$.  From the definitions, one has the \emph{differential} form of the $t$-FDT
\begin{subequations}
\label{eq:FDTC}
\begin{align}
	\label{eq:FDTCt}
	C_{AB}(t)  & =
	 \frac 12 \left [F_{AB}\left(t+\frac {i\beta \hbar}2\right)+
	 F_{AB}\left(t-\frac {i\beta \hbar}2\right) \right ] 
	 	\equiv \hat{\cal{L}}_c  F_{AB}(t) 
		= \cos \left (\frac {\beta \hbar}2 \frac{d}{dt} \right) F_{AB}(t) 
	\\ \nonumber \\
	\label{eq:FDTRt}
	R_{AB}(t) 
	 & 
	  = - \frac{2}{\hbar} \theta(t) \frac{1}{2i}  
	  \left [ F_{AB}\left(t+\frac {i\beta \hbar}2\right)-
	 F_{AB}\left(t-\frac {i\beta \hbar}2\right)\right]
	\equiv
	 - \frac{2}{\hbar} \theta(t)  \hat{\cal{L}}_s  F_{AB}(t)
	 = - \frac{2}{\hbar} \theta(t) \sin \left (\frac {\beta \hbar}2 \frac{d}{dt} \right) F_{AB}(t)  \ .
\end{align}	
\end{subequations}
By taking the classical limit $\beta \hbar \to 0$ one finds the classical version of the FDT
\begin{align}
	\label{eq:class}
	C_{AB}(t) = F_{AB}(t) \ , 
	\quad 
	R_{AB}(t) = - \beta \frac{d C_{AB}(t)}{dt} \ .
\end{align}

A version of the $t$-FDT relating the response $R(t)$ and the fluctuations $C(t)$ was already derived by Pottier and Mauer in Ref.\cite{Pottier2001Quantum}. Here we focus on the relation on $F(t)$ as a function of the fluctuation $C(t)$ or the response $R(t)$. In Section \ref{app:DIFFE} we show how to retrieve the Pottier-Mauer results and to verity our results via application of the differential operators to properly regularized blurring functions.

\subsection{From $C(t)$ and $R(t)$ to $F(t)$}

We now write the inverse of the differential $t-$FDT \eqref{eq:FDTC} as convolutions. We start by considering the aFT of \eqref{eq:fdtC}
\begin{align}
\label{eq:FregC}
	F_{AB}(t) & =  \int_{-\infty}^{\infty} \frac {d\omega}{2 \pi}  e^{-i\omega t}\frac{C_{AB} (\omega) }{\cosh(\beta  \hbar\omega/2)} 
	= \int_{-\infty}^{\infty} {\rm d} t' C_{AB}(t') \,  g_{\Omega}(t-t')  \ .
\end{align}
The function $g_\Omega(t)$ is the anti-Fourier transform of $1/\cosh(\beta \hbar \omega/2)$ and it can be written as
\begin{align}
	\label{eq:GOmega}
	g_{\Omega}(t) 
	= \lim_{\epsilon\to0} g_{\Omega, \epsilon}(t) 
	= \frac {\Omega}{\pi} \frac 1{\cosh(\Omega t) } \ .
\end{align}
where we have defined the regularized anti-Fourier transform 
\begin{equation}
	g_{\Omega, \epsilon}(t) \equiv 
	 \int_{-\infty}^\infty \frac{ e^{-i\omega t - \epsilon |\omega|}}{\cosh(\omega \beta \hbar /2) }
	  \frac{d \omega}{2 \pi } \ .
\end{equation}
The latter can be computed explicitly in terms of the digamma function $\psi(x)$, a special function that obeys the two fundamental properties
\begin{eqnarray}
& & \psi(x+1)-\psi(x) = \frac 1 x  \label{primera} \ ,\\  
& &\psi(1-x) - \psi(x) = \pi \cot(\pi x) \label{segunda} \ .
\end{eqnarray}
By explicit computation we obtain
\begin{eqnarray}
	g_{\Omega,\epsilon}(t)  
	&=& 
	\label{eq:GO1}
	 \frac 1{2 \pi \beta\hbar}
	 \left [\psi \left(\frac{\epsilon-it}{2\beta\hbar} + \frac 34 \right)
	 -\psi \left(\frac{\epsilon-it}{2\beta\hbar} + \frac 14 \right) 
	 	 +\psi \left(\frac{\epsilon+it}{2\beta\hbar} + \frac 34 \right) 	
	 	 - \psi  \left(\frac{\epsilon+it}{2\beta\hbar} + \frac 14 \right)
	 \right ] \\
	 	 	&=& 
	 \label{eq:GO2}
	 \frac 1{2 \pi \beta \hbar}
	 \left [ + \psi \left(\frac{\beta \hbar /2-\epsilon+it}{2\beta\hbar}  \right)
	 	-\psi \left(\frac{\epsilon-it + \beta \hbar/2}{2\beta\hbar}\right) 
	 		 +\psi \left(\frac{\beta \hbar/2 - \epsilon-it}{2\beta\hbar}\right)
	 		 - \psi  \left(\frac{\epsilon+it + \beta\hbar /2}{2\beta\hbar} \right)
 	\nonumber
	 \right ] \\
	  \label{eq:GO3}
    & &+ \frac i{2\beta\hbar} \left[ 
    \coth \left(\pi \frac{t + i \beta\hbar/2 - i\epsilon}{2\beta\hbar}  \right ) - 	
   \coth \left(\pi \frac{ t - i\beta \hbar+ i\epsilon\hbar}{2\beta\hbar}  \right)
     \right ] 
     \underset{\epsilon \to 0}{\longrightarrow} \frac{1}{\beta \hbar} \frac 1{\cosh (\pi t/\beta \hbar)} \ ,
\end{eqnarray}
where the last line corresponds to \eqref{eq:GOmega}. \\

We now perform the same manipulations on \eqref{eq:fdtR}. One has
\begin{align}
\begin{split}
    \label{eq:FregR}
	F_{AB}(t)  &=  \int_{-\infty}^{\infty} \frac {d\omega}{2 \pi}  e^{-i\omega t}\frac{R''_{AB}(\omega) }{\sinh(\beta \omega/2)} 
 =  \int_{-\infty}^{\infty} {\rm d} t' \, iR''_{AB}(t')\,   f_\Omega(t-t') \\
 &
 =  \frac{1}{2}  \int_{-\infty}^{\infty} {\rm d} t' \left[ R_{AB}(t') - R_{BA}(-t')  \right]  f_\Omega(t-t')
 \ ,    
\end{split}
\end{align}
If $A=B$ the equation can be simplified to 
\begin{align}
    \label{eq:FregR}
	F_{AA}(t)  
 = \frac 12  \int_{0}^{\infty} {\rm d} t'  R_{AA}(t') \left[ f_\Omega(t-t') - f_\Omega(t+t') \right]
 \ ,
 \end{align}
where we use the definition of
$R''_{AB}(t) = (R_{AB}(t) - R_{BA}(-t))/2i$ and causality, i.e. $R(t')\propto \theta(t').$ 
The function $f_\Omega(t)$ is $\hbar /i$ the anti-Fourier transform of $1/\sinh(\beta \omega/2)$ and it can be written as
\begin{align}
	\label{eq:GOmega_}
	f_{\Omega}(t) 
	= \lim_{\epsilon\to0} f_{\Omega, \epsilon}(t) 
	=-  \frac {1}{\beta} \tanh(\Omega t) \ .
\end{align}
where we have defined the regularized anti-Fourier transform 
\begin{align}
	\label{f}
	f_{\Omega, \epsilon}(t) &\equiv \frac \hbar i  \int_{-\infty}^{\infty} \frac {d\omega}{2 \pi}  \frac{e^{-i\omega t -\epsilon |\omega|} }{\sinh(\beta \hbar \omega/2)}
	= \frac{i}{\pi \beta} \left[ \psi \left(\frac{\epsilon + i t}{\beta  \hbar} + \frac 12 \right)-\psi \left(\frac{\epsilon - i t}{\beta \hbar} + \frac 12 \right)\right] 
	\underset{\epsilon \to 0}{\longrightarrow} -\frac{1}{\beta} {\tanh (\pi t/\beta \hbar)} \ .
\end{align}
By defining the \emph{integrated response}
\begin{align}
	\label{eq:defPsi}
	\frac{d\Psi_{AB}(t)}{dt} \equiv R_{AB}(t) 
	\quad \to \quad 
	\Psi_{AB}(t) = \int_{0}^tR_{AB}(t') \, {\rm d} t' \ .
\end{align}
we have 
\begin{align}
	F_{AB}(t) 
	& = \frac 12 \int_{-\infty}^{\infty} {\rm d} t'  \big[\,\Psi_{AB}'(t')  - \Psi_{BA}'(-t')  \big ] \, f_\Omega(t-t')
	\\
	& = \frac 12 \big[\,\Psi_{AB}(t')  + \Psi_{BA}(-t')  \big ]  f_\Omega(t-t')  \Big|_{-\infty}^\infty
	+ \frac 12  \int_{-\infty}^{\infty} {\rm d} t'  \big[\,\Psi_{AB}(t')  + \Psi_{BA}(-t')  \big ]\,f'_\Omega(t-t')
	\\
	\label{questa}
	& = \frac{\Psi_{AB}(\infty)  + \Psi_{BA}(\infty)}{2\beta} 
	- \frac 1{2\beta} \int_{-\infty}^{\infty} {\rm d} t' \, \,\tilde f_\Omega(t-t') \big[\,\Psi_{AB}(t')  + \Psi_{BA}(-t')  \big ]
\end{align}
where from the first to the second line we have integrated by parts, from the second to the third we have used that $f_\Omega(-\infty) = - f_\Omega(\infty) = 1/\beta$ and that $\Psi_{AB}(\infty)$ is the integrated response over all times, i.e. the \emph{static susceptibility}
\begin{equation}
	\Psi_{AB}(\infty) = \int_{0}^\infty R_{AB}(t') {\rm d} t'\ .
\end{equation}
We have also defined the blurring function $\tilde f_{\Omega}(t)$ from 
\begin{align}
	\tilde f_\Omega(t) \equiv - \beta \frac {d}{dt} f_\Omega(t) =  \Omega \frac 1{\cosh^2(\Omega t)}\ .
\end{align}
By using the integral
\begin{equation}
	\frac 1{2} \int_{0}^{\infty} {\rm d} t' \, \big[\, \tilde f_\Omega(t-t') + \tilde f_\Omega(t+t') \big ]  = 1 \ ,
\end{equation}
one can re-write \eqref{questa} for $B=A$ as
\begin{equation}
    \label{eq_25}
	F_{AA}(t) 
	 = 
	 \frac 1{2\beta} \int_{0}^{\infty} {\rm d} t' \,\left (
	 \Psi_{AA}(t')  - \Psi_{AA}(\infty) \right )
	 \big[\, \tilde f_\Omega(t-t') + \tilde f_\Omega(t+t') \big ] \ .
\end{equation}

\subsection{Action of the differential operators on the blurring functions}
\label{app:DIFFE}

The application of the differential operators $\hat {\mathcal L}_{c/r}$ to the blurring functions inside the integrals in Eqs.\eqref{eq:FregC} and \eqref{eq:FregR} requires a bit of care. In fact, without regularization, it would generate diverging non-integrable functions. Hence one needs to apply the differential operators $\hat {\mathcal L}_{c/r}$ on the regularized ones $f_{\Omega, \epsilon}$ and $g_{\Omega, \epsilon}$ and then to send $\epsilon \to 0$. 
For sake of notations, we introduce the regularized delta function
\begin{align}
	\label{eq:deltaE}
	 \delta_\epsilon(t) \equiv \frac 1 \pi \frac \epsilon{\epsilon^2 + t^2} 
	  \underset {\epsilon \to 0}\longrightarrow \delta(t) \ .
\end{align}

Using the properties of the digamma functions (see below), one finds
\begin{subequations}
	\label{eq:DiffeSmoth}
\begin{align}
	\label{eq:DiffeSmoth1}
	\hat {\mathcal L}_c \,\, g_{\Omega, \epsilon}(t) & = \delta_\epsilon(t)	\ ,
	\\
	\label{eq:DiffeSmoth2}
 \hat{\mathcal L} _s  \,\, f_{\Omega, \epsilon}(t) & =  {-\frac 12}\hbar  \delta_\epsilon(t)
 \ ,	
	\\
	\label{eq:DiffeSmoth3}
	\hat{\mathcal L} _c  \,\, f_{\Omega, \epsilon}(t) 
	& = 
	- \frac {\hbar \Omega}{\pi} \, \text{Re} \left [\coth \left ( \Omega (t - i\epsilon) \right )
	\right ] \ ,
	\\ 
	\label{eq:DiffeSmoth4}
	\hat{\mathcal L} _s  \,\, g_{\Omega, \epsilon}(t) 
	& = - \frac { \Omega}{\pi} \, \text{Re} \left [ \frac 1{\sinh \left ( \Omega (t - i\epsilon) \right )
	} \right ]\ .
\end{align}
\end{subequations}

To check \eqref{eq:DiffeSmoth1} we compute
\begin{align}
	\mathcal L_c \left [   \psi \left(\frac{\epsilon-it}{2\beta} + \frac 34 \right)-\psi \left(\frac{\epsilon-it}{2\beta  \hbar} + \frac 14 \right)\right ] 
	& = \frac 12 \left [
	 \psi \left(\frac{\epsilon-it}{2\beta  \hbar}+ \frac 34 + \frac 14  \right) + \psi \left(\frac{\epsilon-it}{2\beta  \hbar}+ \frac 34 - \frac 14  \right) \right .
	 \\
	 & \quad\quad  \left .
	-  \psi \left(\frac{\epsilon-it}{2\beta \hbar}+ \frac 14 + \frac 14  \right) - \psi \left(\frac{\epsilon-it}{2\beta \hbar}+ \frac 14 - \frac 14  \right)
	\right ] \\
	& = \frac 12 \left [
	 \psi \left(\frac{\epsilon-it}{2\beta \hbar}+ 1  \right)- \psi \left(\frac{\epsilon-it}{2\beta  \hbar}\right)
	\right ]	 =  \frac{\beta  \hbar}{\epsilon-it} \ ,
\end{align}
where we used the properties of the digamma function. From this, together with \eqref{eq:GO1}, \eqref{eq:DiffeSmoth1} follows.

To check \eqref{eq:DiffeSmoth2} we compute
\begin{align}
 {\cal{L}}_s \;  \psi \left(\frac{\epsilon - i t}{\beta\hbar} + \frac 12 \right)
 & =  \frac  12 \left [\psi \left(\frac{\epsilon - i t}{\beta\hbar} \right) - \psi \left(\frac{\epsilon - i t}{\beta\hbar} +1\right) \right ]
 = -  \frac {\beta \hbar}{2i} \frac {1}{\epsilon - it} 
\end{align}
where we have used (\ref{primera}). Pugging it into \eqref{f}, one obtains Eq.\eqref{eq:DiffeSmoth2}.

To obtain \eqref{eq:DiffeSmoth3} we compute
\begin{subequations}
\label{Argh}
	\begin{align}
		\hat {\mathcal L_c} f_{\Omega, \epsilon}(t)&  = 
		 \frac{i}{2 \pi \beta } \left[ \psi \left(\frac{\epsilon + i t}{\beta\hbar} + 1\right) + \psi \left(\frac{\epsilon + i t}{\beta\hbar}\right )
		 -\psi \left(\frac{\epsilon - i t}{\beta\hbar} + 1 \right) -  \psi \left(\frac{\epsilon - i t}{\beta\hbar}\right )
		 \right]
		 \\
		 & = 	 \frac{i}{2 \pi \beta} \left [ \psi \left(\frac{- \epsilon - i t}{\beta\hbar}\right) 
		  + \psi \left(\frac{\epsilon + i t}{\beta\hbar}\right )
		   -\psi \left(\frac{-\epsilon + i t}{\beta\hbar} \right)
		   -  \psi \left(\frac{\epsilon - i t}{\beta\hbar}\right )
		 \right ] 
		 -  \frac{i}{2 \beta\hbar} \left [ \cot \left ( \pi \frac{\epsilon + i t}{\beta\hbar}\right ) - \cot \left ( \pi \frac{\epsilon - i t}{\beta\hbar}\right )
		 \right ] \\
		 & = \text{F}^{smooth}_\epsilon(t) 
		  -  \frac{1}{2 \beta} \left [ \coth \left ( \pi \frac{t - i\epsilon}{\beta \hbar}\right ) + \coth \left ( \pi \frac{t + i \epsilon}{\beta \hbar}\right )
		 \right ] 
		 \underset{\epsilon \to 0}{\longrightarrow} - \frac{1}\beta \text{P.v.} \coth (\Omega t) 
		 \ ,
	\end{align}
\end{subequations}
where on the right hand side we have defined a smooth function $  \text{F}^{smooth}_\epsilon(t)$ that vanishes for $\epsilon \to 0$.  With similar steps we obtain
\begin{align}
	\hat{\mathcal L}_s g_{\Omega, \epsilon} (t) = 
	& = \text{G}^{smooth}_\epsilon(t) 
		  -  \frac{1}{2  \beta \hbar } \left [ \frac 1{\sinh \left ( \pi \frac{t - i\epsilon}{\beta \hbar}\right )} + \frac 1{\sinh \left ( \pi \frac{t + i\epsilon}{\beta \hbar}\right )} 		 \right ] 
		 \underset{\epsilon \to 0}{\longrightarrow} - \frac{1}{\beta \hbar} \text{P.v.} \frac 1{\sinh (\Omega t) }
		 \ ,
\end{align}
where on the right hand side we have defined a smooth function $  \text{G}^{smooth}_\epsilon(t)$ that vanishes for $\epsilon \to 0$.

\subsection{Relations between $C(t)$ and $R(t)$ }
Therefore, combining the differential equations \eqref{eq:FDTC} together with the regularized Eqs.\eqref{eq:FregC}-\eqref{eq:FregR} we have
\begin{align}
    C_{AB}(t) & =  \lim_{\epsilon \to 0}\int_{-\infty}^{\infty} {\rm d} t' \, iR''_{AB}(t') \,  {\mathcal L}_c \, f_{\Omega, \epsilon}(t-t') 
 = - \frac {1}{\beta}   \text{P.v.} \int_{-\infty}^{\infty} {\rm d} t' \, iR''_{AB}(t')\, 
    \, \coth \left ( \Omega (t-t') \right )
\end{align}
that corresponds to Eq.(2.12) in Ref.\cite{Pottier2001Quantum}.

\subsection{The classical limit $\beta \hbar \to 0$}
The blurring functions $g_\Omega(t)$ and $f_\Omega(t)$ are peaked functions of hight $\propto \Omega$ and width $1/\Omega$ and $1/2\Omega$ respectively. Hence in the classical limit $\Omega \to \infty$ they tend to become delta functions, i.e.
\begin{subequations}
\begin{align}
	\lim_{\Omega \to \infty} g_\Omega(t) &= \delta(t) \\
	\lim_{\Omega \to \infty} f_\Omega(t) &= 2 \delta(t) \ ,
\end{align}	
\end{subequations}
and we immediately retrieve the classical limit in \eqref{eq:class}. 

\section{Examples of the bounds in the frequency domain}
\label{app:frequency}
For simplicity, from now on, we shall concentrate on two-point functions associated with one single operator $\langle A(t) A\rangle$, and neglect the sub-indices, i.e. $C(t)=C_{AA}(t)$ etc.\\
The structure of the FDT bounds the rates of exponentially decreasing correlation functions. In this case, one can reason both in the frequency or in the time domain. Here we discuss two models, whose $\text{Im}R(\omega)$ and $C(\omega)$ correspond to Lorentzians. These are often described in the literature as phenomenological models for transport \cite{Allen2006Conceptual, forster}.

\subsection{Lorentzian dissipation $\text{Im}R(\omega)$}
\label{sec:Jorge}

We start by considering a model and an observable $A$, whose frequency response function has an imaginary part as
\begin{equation}
\label{eq:RaJorge}
	\text{Im} R(\omega)= D  \frac{\omega}{\omega^2 + a^2} \ .
\end{equation}
 This form of dissipation is a very common phenomenological description of transport properties. It corresponds to the Drude model for the conductivity in metals \cite{Allen2006Conceptual} or to the magnetization-magnetization response in a spin diffusion problem \cite{forster}. The response function in time decays exponentially with rate $a>0$, i.e. 
 \begin{equation}
    \label{eq:CaJorgeB_}
	R(t) = D \,  e^{- a t} \, \theta(t) \ .
\end{equation}
The FDT in \eqref{eq:fdt} gives
\begin{equation}
	\label{eq:CaJorge}
	C(\omega) =D \hbar \, \frac{\omega \, \coth (\beta \hbar \omega /2)}{\omega^2 + a^2} \ ,
	\quad 
	F(\omega)  = D \hbar \, \frac{\omega}{\omega^2 + a^2} \frac 1{\sinh(\beta\hbar \omega/2)} \ .
\end{equation}
While $\text{Im}R(\omega)$ admits simple poles only in $\pm  ia $, due to the thermal factors the fluctuations $C(\omega)$ and $F(\omega)$ have poles also in $\pm i 2 n \, \Omega$ with $n=1, 2, \dots$, the so-called bosonic Matsubara frequencies. Furthermore, $C(\omega)$ admits zeros on $\pm i\,  \left( 1 + 2n \right ) \Omega$  with $n=0, 1, 2, \dots$. 
By properly computing the anti-Fourier transform one has
\begin{subequations}
\label{eq:CJorge}
\begin{align}
    F(t)  = 
    \frac {D \hbar} {2 \pi}\,  \int_{-\infty}^{\infty} {d\omega} e^{-i \omega t}\,\,  \frac{\omega }{ \sinh(\beta \hbar \omega /2) (\omega^2 + a^2)}
   &  =	
	 D \left [\frac {  \hbar}{2 \sin(\beta \hbar  a/2)}  e^{- a t}- \frac{8 \Omega}{\beta}	 \sum_{n=0}^\infty \frac{ (-1)^{n} n e^{- 2n\Omega t}}{a^2-(2n\Omega)^2} \right ] \ ,
	 \\
	C(t)
	\label{eq:CJorgeInt}
	=\frac {D \hbar} {2 \pi}\,  \int_{-\infty}^{\infty} {d\omega} e^{-i \omega t}\,\,  \frac{\omega \, \coth (\beta \hbar \omega /2)}{\omega^2 + a^2}
	& = D   \left [  
		\frac  \hbar2 \cot \left ( \frac{\beta\hbar  a }2 \right ) e^{-a t} - \frac {8 \Omega}{\beta }\sum_{n=0}^\infty \frac{n e^{- 2n \Omega t}}{a^2-(2n\Omega)^2} 
		\right ] \ ,
\end{align}
\end{subequations}
where we have solved the complex integral with the contour in the lower half plane (such that $e^{-i (-iz) t}$ decays) and we selected the negative simple poles. The large-time behavior of $F(t)$ and $C(t)$ is determined by the smallest poles on the imaginary axis, therefore as soon as $a>2\Omega$ exceeds the first Matsubara imaginary frequency, the rate $2\Omega$ dominates. This can be summarized as follows
\begin{align}
\label{eq:CaJorgeB}
	C(t) \sim F(t) \sim e^{- t / \tau}
	\quad \text{with} \quad
	\frac 1 \tau = 
	\begin{cases}
 	a \,\,\, \quad \text{for}\quad a < 2\Omega \\
   2\Omega \quad \text{for}\quad a > 2\Omega 
 \end{cases}\quad \leq2 \Omega \ .
\end{align}
By assuming an exponentially decaying response \eqref{eq:CaJorgeB_}, we have obtained a bound on the rate of the exponential decay of the fluctuations function \eqref{eq:CaJorgeB}.

\subsection{Lorentzian fluctuations $C(\omega)$}
\label{sec:Laura}

We now consider a model for which - for some observable $A$ - the fluctuations in frequency are
\begin{equation}
	C(\omega) = B\,  \frac{2 a}{\omega^2 + a^2} \ ,
\end{equation}
whose anti-Fourier transform yield an exponentially decreasing function with a rate $a$, i.e. 
\begin{equation}
\label{eq:CaLAura}
	C(t) = B\,  e^{- a |t|} \ .
\end{equation}
Note that this $C(\omega)$ and the dissipation in the previous section [cf. \eqref{eq:RaJorge}] perfectly satisfy the classical FDT with $D = 2 a B$.
From the quantum FDT in \eqref{eq:fdt}, one has
\begin{equation}
	\label{eq:RaLAura}
	F(\omega)  = B \frac{2 a}{\omega^2 +  a^2} \frac 1{\cosh(\beta\hbar \omega/2)} \ ,
	\quad 
	\text{Im} R(\omega) =B \frac {2  a} \hbar \, \frac{\tanh (\beta \hbar \omega /2)}{\omega^2 +  a^2} \ \ .
\end{equation}
Here, while $C(\omega)$ admits simple poles only in $\pm i a$,  the other $F(\omega)$ and $\text{Im} R(\omega)$ have poles also in $\pm i \Omega(1 + 2n)$ for $n=0, 1, \dots$, the so-called fermionic Matsubara frequencies. The time-dependent functions read
\begin{subequations}
\begin{align}
	R(t)
	\label{eq:RLauraInt}
	& = \frac B \hbar \frac {2 a  i} {\hbar \pi}\, \theta(t)  \int_{-\infty}^{\infty} {d\omega} e^{-i \omega t}\,\, \frac{\tanh (\beta \hbar \omega /2)}{\omega^2 + a^2}
	= \frac {2 B} {\hbar}\, \theta(t)  \left [  \tan \left ( \frac{\beta\hbar  a}2 \right ) e^{-a t} 
	+  \frac {2\Omega} \pi \, {2 a}\sum_{n=0}^\infty \frac{ e^{-(1+ 2n)\Omega t}}{a^2- \Omega^2(1+ 2n)^2}
	\right ] \ ,
	\\
	F(t) & = 
    B\,  \int_{-\infty}^{\infty} \frac{d\omega}{2 \pi} e^{-i \omega t}\,\,  \frac{ 2 a}{ \cosh(\beta \hbar \omega /2) (\omega^2 + a^2)}
	\label{eq:RLauraRes}
	 = B \left [ \frac{ e^{-a t }}{\cos(\beta \hbar a/2)} + \frac {2\Omega} \pi \, {2 a}\sum_{n=0}^\infty \frac{(-1)^n e^{-(1+ 2n)\Omega t}}{a^2-\Omega^2(1+ 2n)^2}
	\right ] \ .
\end{align}
\end{subequations}
We have solved the complex integral with the contour in the lower half plane (such that $e^{-i (-iz) t}$ decays) and we selected the negative simple poles. As before, the large-time behavior of $F(t)$ and $R(t)$ is determined by the smallest poles on the imaginary axis, therefore as soon as $a>\Omega$ exceeds the first fermionic Matsubara imaginary frequency, the rate $\Omega$ dominates. This can be summarized as follows
\begin{align}
\label{eq:CaLauraB}
	R(t) \sim F(t) \sim e^{- t / \tau}
	\quad \text{with} \quad
	\frac 1 \tau = 
	\begin{cases}
 	a \quad \text{for}\quad a <  \Omega \\
 	\Omega \quad \text{for}\quad a > \Omega
 \end{cases} \quad \leq \Omega \ .
\end{align}
By assuming an exponentially decaying fluctuation \eqref{eq:CaLAura}, we have obtained a bound on the rate of the exponential decay of the response function \eqref{eq:CaLauraB}. 

\section{$t-$FDT on correlations functions that decrease exponentially}
\label{app:decay}

\subsection{Starting from $C(t)$}
\label{sec:tFDT_C}
We now evaluate the details of the blurring starting from a correlation function
\begin{equation}
C(t) = D e^{- a |t|} \qquad \text{for} \qquad t\gg t_d \ .
\end{equation}
Repeating the same steps on the $t$-FDT from Eq.(11b) of the main text, we have:
\begin{equation}
\label{eq:F_C}
\begin{array}{ll}
\displaystyle
F(t) = \frac{\Omega}{\pi} \int_{-\infty}^{\infty} {\rm d} t' C(t')  \frac{1}{\cosh \Omega(t-t') } 
\\ \vspace{-0.2cm} \\
\displaystyle
    = \frac{2\Omega}{\pi} \left[ \int_{-\infty}^{t-T} {\rm d} t' C(t')  e^{-\Omega (t-t')} 
    +\frac D2 e^{- a t} \int_{-T}^{\infty} {\rm d} t' e^{-a t'}  \frac{1}{\cosh \Omega t'}  \right] + \dots
 = F_1(t)+F_2(t) + \dots
\end{array}
\end{equation}
Let us re-write the first term as
\begin{align}
F_1(t) & = \frac{2\Omega}{\pi}   
\left ( e^{-\Omega t}\int_{-\infty}^{t_d} {\rm d} t'  C(t') e^{\Omega t'} + 
D\int_{t_d}^{t-T} e^{(- a + \Omega) t'}
\right )
\\
& =  \frac{2\Omega}{\pi}   
\left ( e^{-\Omega t}\int_{-\infty}^{t_d} {\rm d} t'  C(t') e^{\Omega t'} + 
D \frac{e^{-a t} e^{(a-\Omega)T}}{\Omega -a} - D \frac{e^{- \Omega t} e^{(\Omega -a) t_d}}{\Omega - a} 
\right ) \ .
\end{align}
Plugging this back into \eqref{eq:F_C}, all together we have 
\begin{equation}
	F(t) \simeq  c_a e^{- a t} + c_\Omega e^{- \Omega t} + \dots \ ,
\end{equation}
with
\begin{subequations}
	\begin{align}
	    \label{Ca}
		c_a & = 
		D \frac{2\Omega}{\pi}
		\left( \frac{e^{(a-\Omega)T}}{\Omega - a} + \frac 12 \int_{-T}^{\infty} {\rm d} t' e^{-a t'}  \frac{1}{\cosh \Omega t'} 
		\right) = \frac D{\cos(\pi a / 2 \Omega)} \ .
		\\
		c_\Omega & = 
		\frac{2\Omega}{\pi}   
	\left (
	\int_{-\infty}^{t_d} {\rm d} t'  C(t') e^{\Omega t'} 
	- D \frac{ e^{(\Omega -a) t_d}}{\Omega - a} 
	\right ) \ .
	\end{align}
\end{subequations}
These constants can be written in an compact way. 
Let us introduce a cutoff $\Lambda$ (that we will send to infinity) and compute
\begin{align}
\frac{2\Omega}{\pi} \frac{1}{\Omega-a}   
e^{(a -\Omega)T} 
& = \frac{2\Omega}{\pi}  
\left (
\int_{-\Lambda}^{-T} e^{-(a-\Omega)u} {\rm d}u
+ \frac{e^{(a-\Omega)\Lambda}}{a-\Omega}
\right )
\simeq  \frac{\Omega}{\pi}  \int_{-\Lambda}^{-T} e^{-a u} \frac{1}{\cosh\Omega u} {\rm d}u
+ \frac{2\Omega}{\pi}\frac{e^{(a-\Omega)\Lambda}}{a-\Omega} \ .
\end{align}
We substitute it back into \eqref{Ca} and take the limit $\Lambda \to \infty$
\begin{equation}
	c_a=  \frac{\Omega}{\pi}
	\lim_{\Lambda \to \infty}
	\left (\int_{-\Lambda}^{\infty} e^{-a u} \frac{1}{\cosh\Omega u} {\rm d}u + 2\frac{e^{(a-\Omega)\Lambda}}{a-\Omega}
	\right )
	= \frac{1}{\cos(\pi a/ 2 \Omega)}\ .
\end{equation}
On the other hand for $a>\Omega$ we have 
\begin{align}
	c_\Omega =  \frac{2\Omega}{ \pi} 
    \left ( \int_{-\infty}^{t_d}{\rm d} t'  C(t') e^{\Omega t'}  
    + D \int_{t_d}^\infty {\rm d} t' e^{(-a + \Omega) t'}
    \right )
    = \frac{2\Omega}{ \pi}  \int_{-\infty}^{\infty} C(t') e^{\Omega t'}
    \ .
\end{align}
Notice that by using $C(\omega)= D\frac{2a}{\omega^2 + a^2}$ we retrieve the coefficients of the toy model in Section \ref{sec:Laura}, see also \eqref{eq:RLauraRes}. 
One can repeat the calculation with the other terms stemming from the expansion of $\cosh x$ and obtain that for $a>2 n \Omega$ and obtain further contributions to $F(t)$ in the form 
\begin{equation}
    (-1)^{n+1} c_{(1+2n)\Omega} \,\,e^{- (1 + 2 n) \Omega  t}
\quad
\text{with}
\quad
    c_{(1+2n)\Omega} =  \frac{\Omega}{ \pi}  \int_{-\infty}^{\infty} C(t') e^{(1+2n)\Omega t'}
\end{equation}

\subsection{Starting from $R(t)$}
\label{sec:tFDT_R}

We now evaluate the details of the blurring starting from an exponentially decreasing response function $R(t)=De^{-a t}$ for $t\gg t_d$ and some positive rate $a$. 
Its integrated response is
\begin{equation}
\label{eq:PSiExp}
\Psi(\infty) - \Psi(t) = \frac{D}{a} e^{- a t} \qquad \text{for}\qquad t\gg t_d \ ,
\end{equation}
as illustrated in Fig.2a of the main text.
Here $t_d$ is the time where the exponential decay settles down. In the toy models consider above one has $t_d=0$, while in the models type cft $t_d\simeq 1/\Omega$ .
 Let us write $F(t)$ via the $t$-FDT [cf. Eq.(\ref{eq_25})] as
\begin{equation}
\label{smothFR}
F(t) = \frac{\Omega}{2\beta} \int_0^{\infty} {\rm d} t'  (\Psi(\infty)-\Psi(t')) \left[ \frac{1}{[\cosh \Omega (t-t')]^2} + \frac{1}{[\cosh \Omega(t+t')]^2} \right] = F_+(t) + F_-(t) \ .
\end{equation}
We evaluate term by term as
\begin{align}
    \begin{split}
F_+(t) & =   \frac{\Omega}{2\beta} \int_0^{\infty} {\rm d} t'   \frac{\Psi(\infty)-\Psi(t')}{[\cosh \Omega (t-t')]^2}  
\\
& =\frac{2\Omega}{\beta} e^{-2 \Omega t} \int_0^{t-T} {\rm d} t'  (\Psi(\infty)-\Psi(t')) e^{2\Omega t'} + \frac{\Omega}{2\beta}  e^{-a t}  \int_{-T}^{\infty} {\rm d} t' \frac{D}{a} e^{-a u} \frac{1}{[\cosh \Omega u]^2} 
+ \mathcal O(e^{-4 \Omega t})
    \end{split}
\end{align}
where we have first split the integral $\int_0^\infty {\rm d} t' = \int_0^{t-T} {\rm d} t' + \int_{t-T}^\infty$. Then on the left, taking $T\gg 1/\Omega$,  we made the approximation that for $t' \ll t-T$ one has $[\cosh \Omega (t-t')]^{-2}  \simeq 4 e^{- 2\Omega (t-t')}$. On the right hand side, we have first used \eqref{eq:PSiExp} and then performed the change of variables $t'-t=u$. 
Let us compute the first contribution dividing the integral up to $t_d$, after which we can approximate the solution using \eqref{eq:PSiExp}. One has
\begin{equation}
\begin{array}{ll}
\displaystyle
\frac{2\Omega}{\beta} e^{- 2\Omega t} \int_0^{t_d} {\rm d} t'  (\Psi(\infty)-\Psi(t')) e^{2\Omega t'}  +  \frac{\Omega}{2\beta} e^{-2 \Omega t} \int_{t_d}^{t-T} {\rm d} t'  \frac{D}{a} e^{-a t'} e^{2\Omega t'} 
\\ \vspace{-0.2cm} \\
\displaystyle
=  \frac{2\Omega}{\beta} e^{- 2\Omega t}  \left[ \int_0^{t_d} {\rm d} t'  (\Psi(\infty)-\Psi(t')) e^{2 \Omega t'} +  \frac{De^{- (a- 2\Omega) t_d}}{a (-2\Omega+a)} \right] -  \frac{2\Omega}{\beta} e^{- a t} e^{(a-2\Omega) T} \frac{D}{a (-2\Omega+a)} \ .
\end{array}
\end{equation}
For $F_-(t)$ the blurring function is always centered at $-t\ll0$, hence we can also expand the hyperbolic cosine into
\begin{equation}
F_-(t) =  \frac{2\Omega}{\beta}  e^{-2\Omega t}  \int_0^\infty {\rm d} t'  (\Psi(\infty)-\Psi(t')) e^{-2\Omega t'}  + \mathcal O(e^{- 4 \Omega t})\ .
\end{equation}
Substituting into \eqref{smothFR} we have 
\begin{equation}
    F(t) \sim r_a e^{- a t} + r_{2\Omega} e^{- 2 \Omega t} + \mathcal O(e^{-4 \Omega t})\ ,
\end{equation}
with 
\begin{subequations}
\label{51}
\begin{align}
    \label{51a}
    r_a & = 
    -  D\frac{2\Omega}{\beta}    \frac{e^{(a-2\Omega) T}}{a (-2\Omega+a)}
    + D\frac{\Omega}{2\beta}  \int_{-T}^{\infty} {\rm d} t' \frac{1}{a} e^{-a u} \frac{1}{[\cosh \Omega u]^2} 
    = D\, \frac \hbar 2 \frac 1{\sinh(\pi a /2 \Omega)} \ ,
    \\ 
    \label{51b}
    r_{2\Omega} & = 
    \frac{2 \Omega}{\beta} 
    \left [
    \int_0^{t_d} {\rm d} t'  (\Psi(\infty)-\Psi(t')) e^{2 \Omega t'} +  \frac{De^{- (a- 2\Omega) t_d}}{a (-2\Omega+a)}
    +  \int_0^\infty {\rm d} t'  (\Psi(\infty)-\Psi(t')) e^{-2\Omega t'}
    \right ] \ .
\end{align}
\end{subequations}
These constants can be written in an compact way. 
We introduce a cutoff $\Lambda$ (that we will send to infinity) and compute
\begin{equation}
- \frac{2\Omega}{\beta}  \frac{e^{(a-2\Omega) T}}{a (a-2\Omega)} 
    = \frac{2\Omega}{\beta}  \left ( \int_{-\Lambda}^{-T} \frac{e^{-(a - 2 \Omega) u}}{a}  - \frac{e^{(a-2\Omega)\Lambda}}{a(a-2\Lambda)} \right )
    \simeq 
     \frac{\Omega}{2\beta}   \int_{-\Lambda}^{-T} \frac{1}{a} e^{-a u} \frac{1}{(\cosh{\Omega u})^2}
    - \frac{2\Omega}{\beta} \frac{e^{(a-2\Omega)\Lambda}}{a(a-2\Lambda)} \ .
\end{equation}
We substitute this back into \eqref{51a} and take the limit $\Lambda \to \infty$ leading to
\begin{align}
    r_a = D\frac{\Omega}{2\beta a} 
   \lim_{\Lambda \to \infty} \left (
    \int_{-\infty}^{\infty} {\rm d} t' e^{-a u} \frac{1}{[\cosh \Omega u]^2} 
   - 4  \frac{e^{(a-2\Omega)\Lambda}}{a(a-2\Lambda)} \right )
    = \frac{  \hbar }2 \frac D{\sin( \pi a/2 \Omega )} \ .
\end{align}

On the other hand for $a>2 \Omega$ we have 
\begin{align}
    r_{2\Omega} & = \frac {2 \Omega }\beta 
    \left (
    \int_0^\infty {\rm d} t'  (\Psi(\infty)-\Psi(t')) e^{-2\Omega t'} 
    + 
    \int_0^{t_d} {\rm d} t'  (\Psi(\infty)-\Psi(t')) e^{2\Omega t'} + e^{- (a-2\Omega) t_d} \frac{D}{a (-2\Omega+a)}
    \right )    
    \\
    & = \frac {2 \Omega }\beta 
    \left (
    \int_0^\infty {\rm d} t'  (\Psi(\infty)-\Psi(t')) e^{-2\Omega t'} 
    + 
    \int_0^{\infty} {\rm d} t'  (\Psi(\infty)-\Psi(t')) e^{2\Omega t'} 
    \right )
    \\
    & = 
    \frac {2 }\beta 
    \left (
    -( \Psi  (\infty) - \Psi(t') )e^{-2\Omega t'}\Big|_{0}^\infty - 
     \int_0^\infty {\rm d} t' R(t') e^{-2\Omega t'}
    +( \Psi  (\infty) - \Psi(t') )e^{2\Omega t'}\Big|_{0}^\infty 
    +  \int_0^\infty {\rm d} t' R(t') e^{2\Omega t'}
        \right )
    \\
    & = 
    \frac {2 }\beta 
    \left (
     \int_{-\infty}^\infty {\rm d} t' R(t') e^{2\Omega t'}
    -  \int_{-\infty}^\infty {\rm d} t' R(t') e^{-2\Omega t'}
        \right )     
    =
      \frac {4}\beta 
     \int_{-\infty}^\infty {\rm d} t'  e^{2\Omega t'} \, iR''(t')\ ,
\end{align}
where from the second to the third line one integrates by parts using the definition of the integrated response $\psi'(t') = R(t')$. Notice that by interpreting $ r_{2\Omega} = i \frac 4 \beta \text{Im} \tilde R(- i\, 2 \Omega)$ for $\text{Im} \tilde R(\omega) = \frac{\omega}{\omega^2 + a^2}$ we retrieve the the correct coefficient of the model in Section \ref{sec:Jorge}, see also \eqref{eq:CJorge}. 

One can repeat the calculation with the other terms stemming from the expansion of $\cosh x$ and obtain that for $a>2 n \Omega$ and obtain further contributions to $F(t)$ in the form 
\begin{equation}
    (-1)^{n+1} r_{2n\Omega} e^{- 2 \Omega n t}
\quad
\text{with}
\quad
    r_{2n\Omega} = \frac 4 \beta  \int_{-\infty}^{\infty} {\rm d} t' \, iR''(t')\,  e^{ 2n \Omega t}  \ .
\end{equation}

\section{$t-$FDT on correlations functions that increase exponentially in an interval}
\label{app:growth}

Here we provide the detailed evaluation of the $t$-FDT on correlations functions that grow exponentially in a time-regime $t_d\ll t\ll t_{Ehr}$.

\subsection{Starting from $C(t)$}
Imagine we start from a correlation function that for $t_d \ll t\ll t_{Ehr}$ goes as
\begin{equation}
    C(t) = C_d - D \epsilon e^{at}  = C_d - \tilde C(t)\ ,
\end{equation}
where $C_d$ is a constant depending on $t_d$ and $\tilde C(t)=D e^{at}$ only in that interval and it eventually goes to zero after $t_{Ehr}$.
We now want to study the effect of the $t$-FDT on such $C(t)$. Eq.(11)a of the main text reads
\begin{align}
\label{Ftqui}
F(t) & = C_d  - \frac{\Omega}{\pi} \int_{-\infty}^{\infty} {\rm d} t'    \frac{\tilde{C}(t')}{\cosh \Omega (t-t')}  \nonumber
\\
& =
C_d  - \frac{\Omega}{\pi} \int_{-\infty}^{t-T} {\rm d} t'    \frac{\tilde{C}(t')}{\cosh \Omega (t-t')} 
- \frac{\Omega}{\pi} \int_{t-T}^{t+T} {\rm d} t'    \frac{\tilde{C}(t')}{\cosh \Omega (t-t')} 
- \frac{\Omega}{\pi} \int_{t+T}^\infty {\rm d} t'    \frac{\tilde{C}(t')}{\cosh \Omega (t-t')} 
\\
& = C_d  - \frac{\Omega}{\pi} \left (F_1(t) + F_2(t) + F_3(t) \right ) \nonumber \ .
\end{align}
We evaluate term by term. For $t'<t-T$ we expand the $\cosh(\Omega (t-t'))\sim e^{-\Omega(t-t')}/2$
\begin{align}
    F_1(t) & \simeq
    2 e^{-\Omega t}
    \left (
    \int_0^{t_d} \tilde C(t') e^{\Omega t'} dt' + 
    D\epsilon \int_{t_d}^{t-T}  e^{(\lambda + \Omega)t'} dt'
    \right )
     \\ & = 
     2 
    e^{-\Omega t} \, \int_0^{t_d} \tilde C(t') e^{\Omega t'} dt' + 
    \frac{2 D \epsilon}{\lambda + \Omega} 
    \left (e^{\lambda t}e^{-(\lambda + \Omega)T} - e^{(\lambda + \Omega)t_d}
    \right ) \ .
\end{align}
For the second term, we simply perform a change of variables $u=t-t'$, resulting in
\begin{align}
    F_2(t) & = 
    D \epsilon e^{a t} \int_{-T}^T du \frac{e^{a u}}{\cosh \Omega u} \ .
\end{align}
While for $t'>t+T$ we can expand $\cosh(\Omega (t-t'))\sim e^{\Omega(t-t')}/2$ leading to
\begin{align}
    F_3(t) & \simeq 2 \epsilon D e^{\Omega t }\int_{t-T}^{t_{Ehr}} {\rm d} t' e^{(a-\Omega)t' }
    + 2 e^{\Omega t } \int_{t_{Ehr}}^{\infty } \tilde C(t') e^{-\Omega t'}
    \\
    & = \frac{2\epsilon D}{a-\Omega} 
    \left (
    e^{a t}e^{-(a-\Omega)T}  + e^{\Omega t} e^{(a-\Omega) t_{Ehr} }
    \right )
    + 2 e^{\Omega t } \int_{t_{Ehr}}^{\infty } \tilde C(t') e^{-\Omega t'}
\end{align}
Neglecting the exponentially decaying terms we have
\begin{align}
    F(t) & \simeq
    C_d  - C_a e^{a(t-t_{Ehr})} - C_{\Omega} e^{\Omega(t-t_{Ehr})} + O( e^{3\Omega(t-t_{Ehr})})
\end{align}
with
\begin{subequations}
    \begin{align}
        C_a & = D \frac \Omega \pi 
        \left (
        \frac{ 2 e^{-(a + \Omega)T}}{a+\Omega} 
        + \int_{-T}^T du \frac{e^{a u}}{\cosh \Omega u}
        + \frac{2 e^{- (a-\Omega)T}}{a-\Omega}
        \right )
        = \frac D{\cos(\pi a /2 \Omega)} \ ,
    \\
        C_{\Omega} & = 2 \frac \Omega \pi 
        \left (
        \frac{ D }{a-\Omega} + \int_{t_{Ehr}}^\infty \tilde C(t') e^{-\Omega(t'-t_{Ehr})}
        \right ) \ .
    \end{align}
\end{subequations}
We now use $\int_{t_d}^{t_{Ehr}} {\rm d} t' e^{(a-\Omega)(t'-t_{Ehr})}= \frac 1{a-\Omega}(1-e^{(a-\Omega)(t_d-t_{Ehr})})$. Hence for $a>\Omega$ we can neglect the term $\propto e^{(a-\Omega)(t_d-t_{Ehr})}$ and we find
\begin{equation}
    C_\Omega = \frac{2 \Omega}{\pi} \int_{t_d}^{\infty} \tilde C(t') e^{-\Omega(t'-t_{Ehr})} {\rm d} t' 
    =  \frac{2 \Omega}{\pi} \int_{t_d}^{\infty}  \left (C_d - C(t') \right) e^{-\Omega(t'-t_{Ehr})} {\rm d} t' \ .
\end{equation} 
One can repeat the calculation with the other powers stemming from the $\cosh x$ that lead to contributions to $F(t)$ as 
\begin{align}
   & (-1)^{n+1}C_{\Omega(1+2n)} e^{\Omega(1+2n)(t-t_{Ehr})} 
    \\
    \text{with} \quad 
    C_{\Omega(1+2n)} & =
    \frac{2 \Omega}{\pi} \int_{t_d}^{\infty}  \left (C_d - C(t') \right ) 
    e^{-(1+2n)\Omega(t'-t_{Ehr})} {\rm d} t' \ .
\end{align}

\subsection{Starting from $R(t)$}

Let us consider a response function that grows exponentially in time in an interval as
\begin{equation}
 R(t) = \frac{D \epsilon}{a} e^{ a t} 
 \qquad \text{for} \qquad  t_d \ll t \ll t_{Ehr} \ .
\end{equation}
Hence, its integrated response is
\begin{equation}
 \Psi(t) = \frac{D \epsilon}{a} e^{ a t} -  C  = \tilde{\Psi}(t) -  C  \qquad \text{for} \qquad  t_d \ll t \ll t_{Ehr} \ ,
\end{equation}
with $\tilde{\Psi}(t)$ the exponential part of the integrated response and  $C = e^{-a(t_{Ehr}-t_d)}/a - \Psi(t_d)$ a constant that depends on the interval of validity of the exponential growth and that we will consider $C=-\Psi(t_d)$ since we are interested in an interval with $t_d\ll t_{Ehr}$. We define $\epsilon=e^{-a t_{Ehr}}$. 
We now want to study the effect of the $t$-FDT on such $\Psi(t)$. Eq.(11)b of the main text reads
\begin{align}
\label{Ftqui}
F(t) & = \frac{ \psi(\infty)}{\beta} + \frac C{\beta}  - \frac{\Omega}{2\beta} \int_0^{\infty} {\rm d} t'  \tilde{\psi}(t') \left[ \frac{1}{[\cosh \Omega (t-t')]^2} + \frac{1}{[\cosh \Omega (t+t')]^2} \right] 
\nonumber \\ & 
= \frac{ \psi(\infty)}{\beta} + \frac C {2\beta} -  \frac{\Omega}{2\beta} [ F_+(t) + F_-(t) ]
\end{align}
Exactly as above, one has $F_-(t)=\mathcal O(e^{-2 \Omega t})$. Let us instead focus on $F_+(t)$ and split it in three parts:
\begin{align}
F_+(t) & = 4 \int_0^{t-T} \tilde{\Psi}(t') e^{ 2\Omega (-t+t')} {\rm d} t' + \int_{t-T}^{t+T}  {\rm d} t'  \frac{D \epsilon}{a} e^{ a t'}  \frac{1}{[\cosh \Omega (t-t')]^2}
+ 4\int_{t+T}^{\infty} \tilde{\Psi}(t') e^{-2 \Omega (-t+t')} {\rm d} t'
\\ & = F_+^1(t)+ F_+^2(t)+F_+^3(t) + \dots
\end{align}
and evaluate them one by one as
\begin{align}
F_+^1(t) & = 4 e^{-2\Omega t} \int_0^{t_d} \tilde{\Psi}(t') e^{ 2\Omega t'} {\rm d} t' +4  \int_{t_d}^{t-T}  \frac{D \epsilon}{a} e^{ a t'}   e^{2\Omega (-t+t')}
\\
    & = 4e^{-2\Omega t} \int_0^{t_d} \tilde{\Psi}(t') e^{2 \Omega t'} {\rm d} t' 
    + 4 \frac{D \epsilon}{a} \frac{1}{a+2\Omega} \left[  e^{a t} e^{-(a+2\Omega) T} - e^{-2\Omega t}  e^{(a+2\Omega) t_d}  \right]
\\
& 
\simeq 4\frac{D \epsilon}{a} \frac{1}{a+2\Omega}  e^{a t} e^{-(a+2\Omega) T} =  4\frac{D \epsilon}{a} e^{a t} \int_{-\infty}^{-T} e^{(a+2\Omega)u}
 \simeq  \frac{D \epsilon}{a} e^{a t} \int_{-\infty}^{-T} \frac{e^{a u} }{(\cosh\Omega u)^2} \ .
\end{align}
where from the second to the third line we neglected the terms order $e^{-2 \Omega t}$.
The last term of \eqref{Ftqui} reads
\begin{align}
F_+^3(t)& = 4 \int_{t+T}^{\infty} \tilde{\Psi}(t') e^{ 2\Omega (t-t')} {\rm d} t' 
    =  4 \int_{t+T}^{t_{Ehr}}  \frac{D \epsilon}{a} e^{ a t'} e^{ 2\Omega (t-t')} {\rm d} t' 
+  4 e^{2\Omega t} \int_{t_{Ehr}}^{\infty} \tilde{\psi}(t') e^{-2\Omega t'} {\rm d} t'
\\
& =    4 \frac{D \epsilon}{a} \frac{1}{a-2\Omega} \left[ e^{2\Omega t} e^{(a-2\Omega) t_{Ehr}} - e^{a t} e^{(a-2\Omega) T} \right]
+  4 e^{2\Omega t} \int_{t_{Ehr}}^{\infty} \tilde{\psi}(t') e^{-2\Omega t'} {\rm d} t'    
\end{align}
Therefore, all together, we have that the leading terms contributing to $F(t)$ are
\begin{align}
    F(t) = \frac{\Psi(\infty)+C}{\beta} - 
    R_a e^{a(t-t_{Ehr})} 
    -  R_{2\Omega} e^{2 \Omega (t-t_{Ehr})} 
     - \mathcal O\left (R_{4\Omega} e^{4 \Omega (t-t_{Ehr})} \right ) \ ,
\end{align}
with 
\begin{align}
    R_a & =\frac \Omega {2 \beta} \frac Da
    \left (
    \int_{-\infty}^T \frac{e^{au}}{\cosh^2(\Omega u)} du - 4 \frac{e^{(a-2\Omega T)}}{a-2 \Omega}
    \right ) = \frac{\hbar}{2} \frac D{\sin(\pi a/2 \Omega)}
    \\
    R_{2\Omega} & = \frac {2 \Omega }{\beta} 
    \left (
     \frac{D}{a} \frac{1}{a-2\Omega}   +  e^{2 \Omega t_{Ehr}} \int_{t_{Ehr}}^{\infty} \tilde{\Psi}(t') e^{-2\Omega t'} {\rm d} t'
    \right ) \Big|_{a> 2 \Omega} \nonumber
   \\
   & = \frac \Omega {2 \beta} 
    \int_{t_d}^{\infty} \tilde {\Psi}(t') e^{-2 \Omega (t'-t_{Ehr})} + \mathcal O(e^{(2\Omega-a)(t_{Ehr}-t_d)}) \ .
\end{align}
Since $\tilde \Psi(t) \sim \epsilon e^{\lambda t}$ for $t\lesssim t_{Ehr}$, and afterwards it saturates, it is reasonable to assume that the integral is dominated by times around $t_{Ehr}$ and that $c_1=\mathcal O(1)$. However, as we explain in the main text, this remains in general only an assumption.

By considering the different expansions from the $\cosh x$, we get all the subleading terms for $t\ll t_{Ehr}$, i.e. 
\begin{align}
    (-1)^{n+1} R_{2n\Omega} e^{2 n \Omega (t-t_{Ehr})}
    \quad\text{with}\quad
    R_{2n\Omega} =  \frac {2\Omega }{ \beta}  \int_{t_d}^{\infty}  ({\Psi}(t')-\Psi(t_d)) e^{-2 n \Omega (t'-t_{Ehr})} \ .
\end{align}

\bibliography{bibliography}

\end{document}